\documentclass[ALICE,manyauthors]{cernphprep}
\usepackage[comma,square,numbers,sort&compress]{natbib}
\usepackage{hyperref}
\usepackage{lineno}
\usepackage{xspace}
\usepackage{comment}
\usepackage{rotating}
\usepackage[utf8]{inputenc}
\usepackage{amsmath}
\usepackage[T1]{fontenc}
\begin{document}
%

\newcommand{\pp}           {pp\xspace}
\newcommand{\ppbar}        {\mbox{$\mathrm {p\overline{p}}$}\xspace}
\newcommand{\XeXe}         {\mbox{Xe--Xe}\xspace}
\newcommand{\PbPb}         {\mbox{Pb--Pb}\xspace}
\newcommand{\pA}           {\mbox{pA}\xspace}
\newcommand{\pPb}          {\mbox{p--Pb}\xspace}
\newcommand{\AuAu}         {\mbox{Au--Au}\xspace}
\newcommand{\dAu}          {\mbox{d--Au}\xspace}

\newcommand{\s}            {\ensuremath{\sqrt{s}}\xspace}
\newcommand{\snn}          {\ensuremath{\sqrt{s_{\mathrm{NN}}}}\xspace}
\newcommand{\pt}           {\ensuremath{p_{\rm T}}\xspace}
\newcommand{\meanpt}       {$\langle p_{\mathrm{T}}\rangle$\xspace}
\newcommand{\ycms}         {\ensuremath{y_{\rm CMS}}\xspace}
\newcommand{\ylab}         {\ensuremath{y_{\rm lab}}\xspace}
\newcommand{\etarange}[1]  {\mbox{$\left | \eta \right |~<~#1$}}
\newcommand{\yrange}[1]    {\mbox{$\left | y \right |~<~#1$}}
\newcommand{\dndy}         {\ensuremath{\mathrm{d}N_\mathrm{ch}/\mathrm{d}y}\xspace}
\newcommand{\dndeta}       {\ensuremath{\mathrm{d}N_\mathrm{ch}/\mathrm{d}\eta}\xspace}
\newcommand{\avdndeta}     {\ensuremath{\langle\dndeta\rangle}\xspace}
\newcommand{\dNdy}         {\ensuremath{\mathrm{d}N_\mathrm{ch}/\mathrm{d}y}\xspace}
\newcommand{\Npart}        {\ensuremath{N_\mathrm{part}}\xspace}
\newcommand{\Ncoll}        {\ensuremath{N_\mathrm{coll}}\xspace}
\newcommand{\dEdx}         {\ensuremath{\textrm{d}E/\textrm{d}x}\xspace}
\newcommand{\RpPb}         {\ensuremath{R_{\rm pPb}}\xspace}

\newcommand{\nineH}        {$\sqrt{s}~=~0.9$~Te\kern-.1emV\xspace}
\newcommand{\seven}        {$\sqrt{s}~=~7$~Te\kern-.1emV\xspace}
\newcommand{\twoH}         {$\sqrt{s}~=~0.2$~Te\kern-.1emV\xspace}
\newcommand{\twosevensix}  {$\sqrt{s}~=~2.76$~Te\kern-.1emV\xspace}
\newcommand{\five}         {$\sqrt{s}~=~5.02$~Te\kern-.1emV\xspace}
\newcommand{\twosevensixnn}{$\sqrt{s_{\mathrm{NN}}}~=~2.76$~Te\kern-.1emV\xspace}
\newcommand{\fivenn}       {$\sqrt{s_{\mathrm{NN}}}~=~5.02$~Te\kern-.1emV\xspace}
\newcommand{\LT}           {L{\'e}vy-Tsallis\xspace}
\newcommand{\GeVc}         {Ge\kern-.1emV/$c$\xspace}
\newcommand{\MeVc}         {Me\kern-.1emV/$c$\xspace}
\newcommand{\TeV}          {Te\kern-.1emV\xspace}
\newcommand{\GeV}          {Ge\kern-.1emV\xspace}
\newcommand{\MeV}          {Me\kern-.1emV\xspace}
\newcommand{\GeVmass}      {Ge\kern-.2emV/$c^2$\xspace}
\newcommand{\MeVmass}      {Me\kern-.2emV/$c^2$\xspace}
\newcommand{\lumi}         {\ensuremath{\mathcal{L}}\xspace}

\newcommand{\ITS}          {\rm{ITS}\xspace}
\newcommand{\TOF}          {\rm{TOF}\xspace}
\newcommand{\ZDC}          {\rm{ZDC}\xspace}
\newcommand{\ZDCs}         {\rm{ZDCs}\xspace}
\newcommand{\ZNA}          {\rm{ZNA}\xspace}
\newcommand{\ZNC}          {\rm{ZNC}\xspace}
\newcommand{\SPD}          {\rm{SPD}\xspace}
\newcommand{\SDD}          {\rm{SDD}\xspace}
\newcommand{\SSD}          {\rm{SSD}\xspace}
\newcommand{\TPC}          {\rm{TPC}\xspace}
\newcommand{\TRD}          {\rm{TRD}\xspace}
\newcommand{\VZERO}        {\rm{V0}\xspace}
\newcommand{\VZEROA}       {\rm{V0A}\xspace}
\newcommand{\VZEROC}       {\rm{V0C}\xspace}
\newcommand{\Vdecay} 	   {\ensuremath{V^{0}}\xspace}

\newcommand{\ee}           {\ensuremath{e^{+}e^{-}}} 
\newcommand{\pip}          {\ensuremath{\pi^{+}}\xspace}
\newcommand{\pim}          {\ensuremath{\pi^{-}}\xspace}
\newcommand{\kap}          {\ensuremath{\rm{K}^{+}}\xspace}
\newcommand{\kam}          {\ensuremath{\rm{K}^{-}}\xspace}
\newcommand{\pbar}         {\ensuremath{\rm\overline{p}}\xspace}
\newcommand{\kzero}        {\ensuremath{{\rm K}^{0}_{\rm{S}}}\xspace}
\newcommand{\lmb}          {\ensuremath{\Lambda}\xspace}
\newcommand{\almb}         {\ensuremath{\overline{\Lambda}}\xspace}
\newcommand{\Om}           {\ensuremath{\Omega^-}\xspace}
\newcommand{\Mo}           {\ensuremath{\overline{\Omega}^+}\xspace}
\newcommand{\X}            {\ensuremath{\Xi^-}\xspace}
\newcommand{\Ix}           {\ensuremath{\overline{\Xi}^+}\xspace}
\newcommand{\Xis}          {\ensuremath{\Xi^{\pm}}\xspace}
\newcommand{\Oms}          {\ensuremath{\Omega^{\pm}}\xspace}
\newcommand{\degree}       {\ensuremath{^{\rm o}}\xspace}

\newcommand{\Dzero}     {\ensuremath{\rm D}^{0}\xspace}
\newcommand{\Dplus}     {\ensuremath{\rm D}^{+}\xspace}
\newcommand{\Dstar}     {\ensuremath{\rm D}^{*+}\xspace}
\newcommand{\Dstarwide} {\ensuremath{\rm D}^{*}(2010)^{+}\xspace}
\newcommand{\Dphi}      {\Delta\varphi}
\newcommand{\Deta}      {\Delta\eta}
\newcommand{\ptD}       {\ensuremath{p_{\rm T}}^{\rm D}\xspace}
\newcommand{\ptass}     {\ensuremath{p_{\rm T}}^{\rm assoc}\xspace}

\begin{titlepage}
\PHyear{2019}       
\PHnumber{239}      
\PHdate{25 October}  

\title{Azimuthal correlations of prompt D mesons with charged particles in pp and p--Pb collisions at $\boldsymbol{\sqrt{s_{\rm NN}}} = 5.02$ \TeV } 
\ShortTitle{Prompt D meson-charged particle correlations in \pp, \pPb at $\snn = 5.02$ \TeV}   

\Collaboration{ALICE Collaboration\thanks{See Appendix~\ref{app:collab} for the list of collaboration members}}
\ShortAuthor{ALICE Collaboration} 

\begin{abstract}
The measurement of the azimuthal-correlation function of prompt D mesons with charged particles in \pp collisions at $\s = 5.02$~\TeV and \pPb collisions at $\snn = 5.02$~\TeV with the ALICE detector at the LHC is reported. The $\Dzero$, $\Dplus$, and $\Dstar$ mesons, together with their charge conjugates, were reconstructed at midrapidity in the transverse momentum interval $3 < \pt < 24$~\GeVc and correlated with charged particles having $\pt > 0.3$~\GeVc and pseudorapidity $|\eta| < 0.8$. The properties of the correlation peaks appearing in the near- and away-side regions (for $\Dphi \approx 0$ and $\Dphi \approx \pi$, respectively) were extracted via a fit to the azimuthal correlation functions. The shape of the correlation functions and the near- and away-side peak features are found to be consistent in \pp and \pPb collisions, showing no modifications due to nuclear effects within uncertainties. The results are compared with predictions from Monte Carlo simulations performed with the PYTHIA, POWHEG+PYTHIA, HERWIG, and EPOS 3 event generators.

\end{abstract}
\end{titlepage}

\setcounter{page}{2} 


\section{Introduction}
\label{sec:Intro}
Two-particle angular correlations allow the mechanisms of particle production to be investigated and the event properties of ultra-relativistic hadronic collisions to be studied.
In particular, the azimuthal and pseudorapidity distribution of “associated” charged particles with respect to a “trigger” D meson is sensitive to the charm-quark production, fragmentation, and hadronisation processes in proton--proton (pp) collisions and to their possible modifications in larger collision systems, like proton--nucleus (pA) or \mbox{nucleus--nucleus} (AA)~\cite{Beraudo:2014boa}.
The typical structure of the correlation function, featuring a ``near-side'' (NS) peak at $(\Dphi,\Deta) = (0,0)$ (where $\Dphi$ is the difference between charged-particle and D-meson azimuthal angles $\varphi_{\rm ch} - \varphi_{\rm D}$, and $\Deta$ the difference between their pseudorapidities $\eta_{\rm ch} - \eta_{\rm D}$) and an ``away-side'' (AS) peak at $\Dphi = \pi$ extending over a wide $\Deta$ range, as well as its sensitivity to the different charm-quark production mechanisms, are described in details in~\cite{ALICE:2016clc}.

In this paper, results of azimuthal correlations of prompt D mesons with charged particles at midrapidity in \pp and \pPb collisions at $\snn = 5.02$ \TeV are presented, where ``prompt'' refers to D mesons produced from charm-quark fragmentation, including the decay of excited charmed resonances and excluding D mesons produced from beauty-hadron weak decays.
The study of the near-side correlation peak is strongly connected to the characterisation of charm jets and of their internal structure, in terms of their particle multiplicity and angular profile.
Probing the near-side peak features as a function of the charged-particle transverse momentum ($\pt$), possibly up to values of a few \GeVc, gives not only access to the transverse-momentum distribution of the jet constituents, but can also provide insight into how the jet-momentum fraction not carried by the D meson is shared among the other particles produced by the parton fragmentation, as well as on the correlation between the $\pt$ of these particles and their radial displacement from the jet axis, which is closely related to the width of the near-side correlation peak.
This study provides further and complementary information with respect to the analysis of charm jets reconstructed as a single object through a track-clustering algorithm and tagged by their charm content~\cite{Aad:2011td,Acharya:2019zup,Sirunyan:2016fcs}.

The azimuthal-correlation function of D mesons with charged particles is largely sensitive to the various stages of the D-meson and particle evolution, as hard-parton scattering, parton showering, fragmentation and hadronisation~\cite{Prino:2016cni}. Its description by the available Monte Carlo event generators like PYTHIA~\cite{Sjostrand:2006za,Sjostrand:2007gs}, HERWIG~\cite{Corcella:2000bw,Bellm:2015jjp,Bahr:2008pv}, and EPOS 3~\cite{Werner:2010aa,Drescher:2000ha} or pQCD calculations like POWHEG~\cite{Nason:2004rx,Frixione:2007vw} coupled to event generators handling the parton shower, depends on several features, including the order of the hard-scattering matrix-element calculations (leading order or next-to-leading order), the modelling of the parton shower, the algorithm used for the fragmentation and hadronisation, and the description of the underlying event.
The azimuthal-correlation function of D mesons with charged particles in \pp collisions at $\s = 7$~\TeV measured by ALICE is described within uncertainties by simulations produced using PYTHIA6, PYTHIA8 and POWHEG+PYTHIA6 event generators~\cite{ALICE:2016clc}.
However, more precise and differential measurements are needed to set constraints to models and be sensitive to the differences among their expectations.

The validation of Monte Carlo simulations for angular correlations of heavy-flavour particles in \pp collisions is also useful for interpreting the results in nucleus--nucleus collisions, for which the measurements in \pp collisions are used as reference. The temperature and energy density reached in nucleus--nucleus collisions at LHC energies are large enough to produce a quark--gluon plasma (QGP), a deconfined state of strongly-interacting matter~\cite{Braun-Munzinger:2015hba,Bazavov:2014pvz}. The interaction of heavy quarks (charm and beauty) with the QGP should affect the angular-correlation function~\cite{Nahrgang:2013saa,Cao:2017hpp,Beraudo:2014boa}. First measurements performed at RHIC and the LHC showed modifications of the correlation function in nucleus--nucleus collisions when the trigger was a heavy-flavour particle, where a suppression of the away-side correlation peak and an enhancement of the near-side correlation peak for associated particles with $\pt < 2$ \GeVc was observed~\cite{Adare:2010ud,Adam:2019tql}.
A comparison of the results in nucleus--nucleus collisions to those in pp collisions, along with a successful description by models, would allow the modifications of the correlation function to be related to the in-medium heavy-quark dynamics~\cite{Nahrgang:2013saa,Cao:2015cba,Cao:2016alj}.

In proton--nucleus collisions, several cold-nuclear-matter effects can influence the production, fragmentation and hadronisation of heavy-flavour quarks. They are induced by the presence of a nucleus in the initial state of the collision and, possibly, by the high density of particles in its final state.
The most relevant effect is a modification of the parton distribution functions due to nuclear shadowing~\cite{Eskola:2016oht}, which can consequently affect the heavy-flavour production cross section.
Measurements of the nuclear modification factor of D mesons and of electrons from heavy-flavour hadron decays in $\pPb$ collisions at $\snn = 5.02$ \TeV~\cite{Acharya:2019mno,Adam:2015qda} point towards a small influence of cold-nuclear-matter effects on the heavy-flavour quark production at midrapidity.
Nevertheless, nuclear effects could still affect the fragmentation and hadronisation of heavy quarks. These can be investigated by measuring potential modifications of the shape of the angular correlation between heavy-flavour particles~\cite{Fujii:2013yja} or, more indirectly, between heavy-flavour particles and charged particles.

Additionally, the search and characterisation of collective-like effects in high-multiplicity proton--proton and proton--nucleus collisions are a crucial topic, due to the observation of long-range, ridge-like structures in two-particle angular-correlation functions at RHIC~\cite{Adare:2013piz,Adamczyk:2015xjc} and the LHC~\cite{Abelev:2012ola,ABELEV:2013wsa,Aaboud:2016yar,Chatrchyan:2013nka,Khachatryan:2010gv,Khachatryan:2014jra}, resembling those observed in \PbPb collisions.
The mechanism leading to these structures in small collision systems is not straightforward to identify. Possible explanations include final-state effects due to a hydrodynamic behaviour of the produced particles~\cite{Werner:2010ss,Deng:2011at}, colour-charge exchanges~\cite{Wong:2011qr,Dumitru:2013tja}, initial-state effects, such as gluon saturation as described within the Color-Glass Condensate effective field theory~\cite{Bzdak:2013zma,Dusling:2013qoz}, or gluon bremsstrahlung by a quark-antiquark string~\cite{Arbuzov:2011yr}.
In addition, a positive elliptic-flow coefficient was observed also for heavy-flavour particles, from the analysis of their azimuthal correlations with charged particles, by the ALICE~\cite{Acharya:2018dxy,Acharya:2017tfn,Adam:2015bka}, ATLAS~\cite{ATLAS-CONF-2017-073,ATLAS-CONF-2017-006, Aad:2019aol}, and CMS~\cite{Sirunyan:2018toe,Sirunyan:2018kiz} Collaborations.
This approach generally assumes that the jet-induced correlation peaks do not differ in low- and high-multiplicity collisions, i.e. nuclear effects have the same impact on the heavy-quark fragmentation and hadronisation at different event multiplicities. This assumption can be tested by looking for modifications of the azimuthal-correlation function.

The results presented in this paper significantly improve the precision and extend the kinematic reach, with respect to our previous measurements~\cite{ALICE:2016clc} in both \pp (at a different energy) and minimum bias \pPb collisions.
Correlations with associated particles at higher $\pt$ probe the angular and $\pt$ distribution of the hardest jet fragments, which retain more closely the imprint of the hard-scattering topology.
The properties of the away-side peak are also studied for the first time.
The paper is structured as follows. In Sec.~\ref{sec:ALICE_and_Data}, the ALICE apparatus, its main detectors and the data samples used for the analysis are presented. In Sec.~\ref{sec:Analysis} the procedure adopted for building the azimuthal-correlation functions, correcting them for experimental effects, and extracting physical quantities is described. Section~\ref{sec:Systematics} describes the systematic uncertainties associated to the measurement. The results of the analysis are presented and discussed in Sec.~\ref{sec:Results}. The paper is briefly summarised in Sec.~\ref{sec:Summary}. 

\section{Experimental apparatus and data sample}
\label{sec:ALICE_and_Data}
The ALICE apparatus consists of a central barrel, covering the pseudorapidity region $|\eta|<$ 0.9, a muon spectrometer with $-4 < \eta < -2.5$ coverage, and forward- and backward-pseudorapidity detectors employed for triggering, background rejection, and event characterisation. A complete description of the detector and an overview of its performance are presented in~\cite{Aamodt:2008zz,Abelev:2014ffa}. The central-barrel detectors used in the analysis presented in this paper, employed for charged-particle reconstruction and identification at midrapidity, are the Inner Tracking System (ITS), the Time Projection Chamber (TPC), and the Time-Of-Flight detector (TOF). They are embedded in a large solenoidal magnet that provides a magnetic field of 0.5~T, parallel to the beams.
The ITS consists of six layers of silicon detectors, with the innermost two composed of Silicon Pixel Detectors (SPD). It is used to track charged particles and to reconstruct primary and secondary vertices. The TPC is the main tracking detector of the central barrel. In addition, it performs particle identification via the measurement of the particle specific energy loss (d$E$/d$x$) in the detector gas. Additional information for particle identification is provided by the TOF, via the measurement of the charged-particle flight time from the interaction point to the detector.
The TOF information is also employed to evaluate the starting time of the event~\cite{Adam:2016ilk}, together with the time information provided by the T0 detector, an array of Cherenkov counters located along the beam line, at $+370$ cm and $-70$ cm from the nominal interaction point.

The results reported in this paper were obtained on the data samples collected during the 2016 LHC p--Pb run at $\snn = 5.02$ \TeV and the 2017 LHC pp run at $\s = 5.02$ \TeV, corresponding, after the event selection, to integrated luminosities of $L_{\rm int}$ = (295 $\pm$ 11) $\mu$b$^{-1}$ and $L_{\rm int}$ = (19.3 $\pm$ 0.4) nb$^{-1}$, respectively. The events were selected using a minimum bias (MB) trigger provided by the V0 detector~\cite{Abbas:2013taa}, a system of two arrays of 32 scintillators each, covering the full azimuthal angle in a pseudorapidity range of $2.8 < \eta < 5.1$ (V0A) and $-3.7 < \eta < -1.7$ (V0C). The trigger condition required at least one hit in both the V0A and the V0C scintillator arrays. This trigger is fully efficient for recording collisions in which a D meson is produced at midrapidity~\cite{ALICE:2016clc}.
The V0 time information and the correlation between number of hits and track segments in the SPD were used to reject background events from the interaction of one of the beams with the residual gas in the vacuum tube.
Pile-up events, whose probability was below 1\% (0.5\%) in pp collisions ($\pPb$ collisions), were rejected with almost 100\% efficiency by using an algorithm based on track segments, reconstructed with the SPD, to detect multiple primary vertices. The remaining undetected pile-up events are a negligible fraction of the analysed sample.
In order to obtain a uniform acceptance of the detectors, only events with a reconstructed primary vertex within $\pm$10 cm from the centre of the detector along the beam line were considered for both pp and $\pPb$ collisions. In $\pPb$ collisions, the $\snn = 5.02$ \TeV energy was obtained by delivering proton and lead beams with energies of 4 TeV and 1.58 TeV per nucleon, respectively. Therefore, the proton--nucleus center-of-mass frame was shifted in rapidity by $\Delta y_{\rm{NN}}$ = 0.465 in the proton direction with respect to the laboratory frame.
The azimuthal correlations between D mesons and charged particles in $\pPb$ collisions were studied as a function of the collision centrality. The centrality estimator is based on the energy deposited in the zero-degree neutron calorimeter in the Pb-going direction (ZNA). The procedure used to define the centrality classes and to determine the average number of binary nucleon--nucleon collisions for each class is described in~\cite{Adam:2014qja}.

Some of the corrections for the azimuthal-correlation functions described in Sec.~\ref{sec:Analysis} were evaluated by exploiting Monte Carlo simulations, which included a detailed description of the apparatus geometry and of the detector response, using the GEANT3 package~\cite{Brun:1082634}, as well as the luminous region distribution during the pp and $\pPb$ collision runs. For the evaluation of the charged-particle reconstruction efficiency, pp collisions were simulated with the PYTHIA8 event generator~\cite{Sjostrand:2007gs} with Monash-2013 tune~\cite{Skands:2014pea}, while $\pPb$ collisions were simulated using the HIJING 1.36 event generator~\cite{PhysRevD.44.3501} in order to describe the charged-particle multiplicity and detector occupancy observed in data~\cite{ALICE:2012xs}.
For the corrections requiring the presence of a D meson in the event, enriched Monte Carlo samples were used, obtained by generating \pp collisions containing a ${\rm c}\overline{\rm c}$ or ${\rm b}\overline{\rm b}$ pair in the rapidity range $[-1.5,1.5]$, employing PYTHIA 6.4.21 with Perugia-2011 tune. For \pPb collisions, an underlying event generated with HIJING 1.36, was superimposed to each heavy-quark enhanced PYTHIA event.

\section{Data analysis}
\label{sec:Analysis}

The analysis largely follows the procedure described in detail in~\cite{ALICE:2016clc}. It consists of three main parts: (i) reconstruction and selection of D mesons and primary charged particles (see~\cite{ALICE-PUBLIC-2017-005} for the definition of primary particle); (ii) construction of the azimuthal-correlation function and corrections for detector-related effects, secondary particle contamination, and beauty feed-down contribution; (iii) extraction of correlation properties via fits to the average D-meson azimuthal-correlation functions with charged particles.

\subsection{Selection of D mesons and primary charged particles}
The analysis procedure begins with the reconstruction of D mesons ($\Dzero$, $\Dstarwide$, and $\Dplus$ and their charge conjugates), defined as ``trigger" particles, and primary charged particles, considered as ``associated" particles. The D mesons are reconstructed from the following hadronic decay channels: ${\Dzero \to {\rm K}^{-}\pi^{+}}$ (BR~=~3.89 $\pm$ 0.04\%), ${\Dplus \to {\rm K}^{-}\pi^{+}\pi^{+}}$ (BR~=~8.98 $\pm$ 0.28\%), and ${\Dstar \to \Dzero\pi^{+}} \to {\rm K}^{-}\pi^{+}\pi^{+}$ (BR~=~2.63 $\pm$ 0.03\%)~\cite{Tanabashi:2018oca} in the transverse-momentum interval $3 < \pt < 24$~\GeVc. The D-meson selection strategy, described in detail in~\cite{Acharya:2019mgn, Acharya:2019mno}, exploits the displaced topology of the decay and utilises the particle identification capabilities of the TPC and TOF to select on the D-meson decay particles. A dedicated optimisation on the selection variables was done, where the selections were tightened to increase the signal-to-background ratio of the D-meson invariant mass peaks. A gain up to a factor 5 at low $\ptD$ was obtained with respect to the selection defined in~\cite{Acharya:2019mgn, Acharya:2019mno}, at the expenses of a reduction of the raw yield. This allowed reducing the impact of the D-meson combinatorial background, whose subtraction induces the largest source of statistical uncertainty on the correlation functions.
With the adopted candidate selection, the D-meson reconstruction efficiency is of the order of few percent for $\ptD = 3$~\GeVc and increases up to 35\% (50\%) for $\ptD = 24$~\GeVc in case of $\Dzero$ and $\Dplus$ ($\Dstar$) both in pp and in \pPb collisions.

\begin{figure}[tb]
    \begin{center}

    \includegraphics[width = 0.99\textwidth]{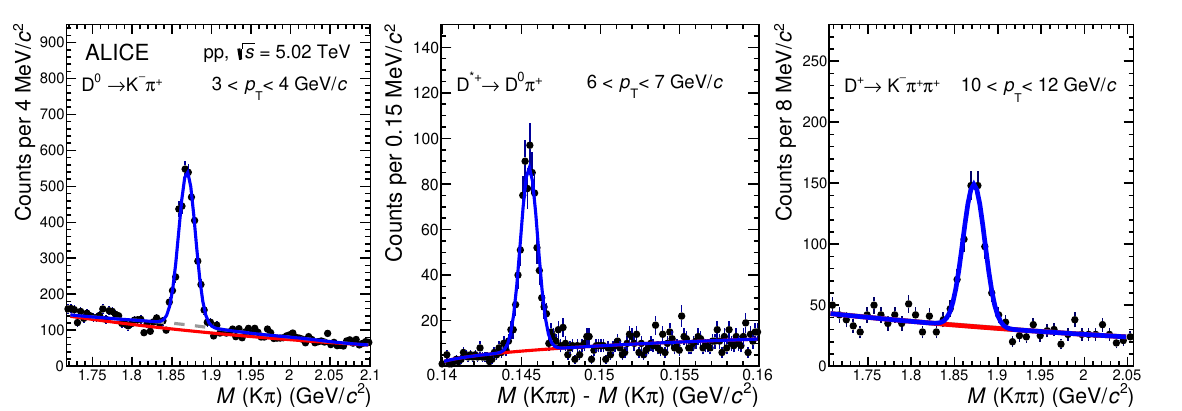}\\
    \includegraphics[width = 0.99\textwidth]{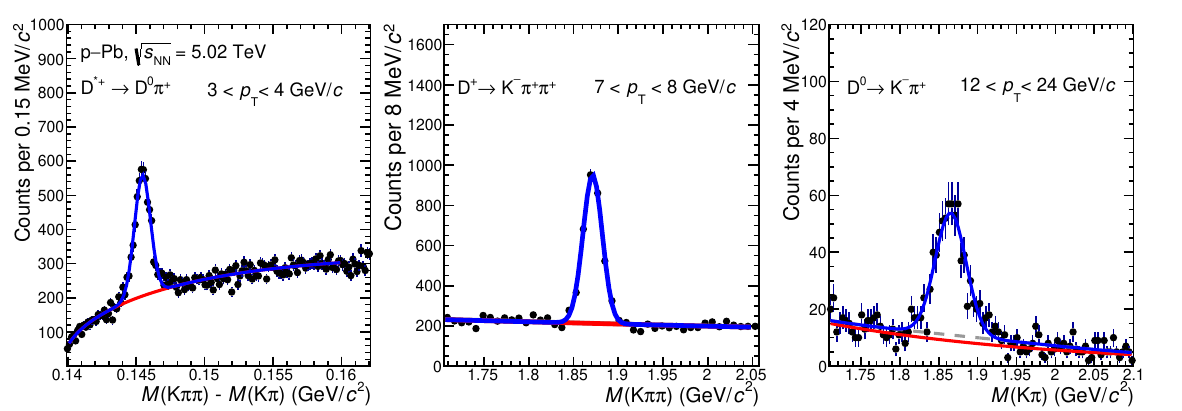}

    \caption{Invariant mass (mass-difference) distributions of $\Dzero$, $\Dplus$ ($\Dstar$), and charge conjugates, candidates in three $\ptD$ intervals for pp collisions at $\s$ = 5.02 TeV (top row) and \pPb collisions at $\snn$ = 5.02 TeV (bottom row). The curves show the fit functions applied to the distributions. For the $\Dzero$, the dashed line represents the combinatorial background including the contribution of reflection candidates (see~\cite{Acharya:2019mgn}). 
    }
    \label{fig:InvMass}
    \end{center}
\end{figure}

The D-meson raw yields were extracted from fits applied to the invariant mass ($M$) distributions of $\Dzero$ and $\Dplus$ candidates, and to the distribution of the mass difference $\Delta M = M(\rm{K}\pi\pi) - M(\rm{K}\pi)$ for $\Dstar$ candidates, for several sub-ranges in the interval 3 $< \pt <$ 24~\GeVc. The fit function was composed of two terms, one for the signal and one for the background. The signal was described by a Gaussian, while the background was modelled by an exponential term for $\Dzero$ and $\Dplus$ mesons, and by a threshold function multiplied by an exponential for the $\Dstar$ meson, as detailed in~\cite{ALICE:2016clc}.
Examples of the invariant mass distributions in \pp and in \pPb collision systems are shown in Fig.~\ref{fig:InvMass} for $\Dzero$, $\Dplus$, and $\Dstar$ mesons in different $\pt$ intervals.

Associated particles are defined as charged primary particles with $\ptass >$ 0.3~\GeVc and with pseudorapidity $|\eta| <$ 0.8. As additional requirement, for this study only pions, kaons, protons, electrons and muons are considered as associated particles. The associated-particle sample does not include the decay products of the trigger D meson.
Reconstructed charged-particle tracks with at least 70 space points out of 159 in the TPC, 2 out of 6 in the ITS, and a $\chi^2$/ndf of the momentum fit in the TPC smaller than 2 were considered.
The contamination of non-primary particles
was largely suppressed by requiring the distance of closest approach (DCA) of the track to the primary vertex to be less than 1 cm in the transverse ($xy$) plane and along the beam line ($z$-direction). This selection identifies primary particles with a purity varying from 95\% to 99\% (increasing with $\ptass$) and rejects a negligible amount of primary particles. In particular, less than 1\% of the primary particles originating from decays of heavy-flavour hadrons are discarded.
For the $\Dzero$ mesons produced in $\Dstar \to \Dzero\pi^{+}$ decays, the low-$\pt$ pion accompanying the $\Dzero$ was removed from the sample of associated particles by rejecting tracks that, combined with the $\Dzero$, yielded a $\Delta M$ consistent within 3$\sigma$ with the $\Dstar$ mass peak. It was verified with Monte Carlo simulations that this selection rejects more than 99\% of the pions from $\Dstar$ decays in all D-meson $\pt$ intervals considered and has an efficiency larger than 99\% for primary particles with $\ptass >$ 0.3~\GeVc.
The selection criteria described above provided an average track reconstruction efficiency for charged particles with $\ptass > 0.3$~\GeVc of about 83\% (82\%) in pp (\pPb) collisions in the pseudorapidity interval $|\eta| <$ 0.8, with an increasing trend as a function of $\ptass$ up to $\approx1$~\GeVc, followed by saturation at about 90\%. As the track reconstruction efficiency has a sudden drop below $\approx0.3$~\GeVc, caused by the TPC requirements in the track selection, this transverse momentum value was chosen as the minimum $\ptass$ for the analysis.

\subsection{Evaluation and correction of the azimuthal-correlation functions}
Selected D-meson candidates with an invariant mass in the range $|M - \mu|~<~2\sigma$ (peak region), where $\mu$ and $\sigma$ denote the mean and width of the Gaussian term of the invariant mass fit function, were correlated to the primary charged particles selected in the same event. A two-dimensional angular-correlation function $C(\Dphi, \Delta\eta)_{\rm{peak}}$ was evaluated by computing the difference of the azimuthal angle and the pseudorapidity of each pair. The azimuthal-correlation functions were studied in four D-meson $\pt$ intervals: $3 < \ptD < 5$~\GeVc, $5 < \ptD < 8$~\GeVc, $8 <\ptD< 16$~\GeVc, and $16 < \ptD < 24$~\GeVc and in the following $\pt$ ranges of the associated tracks: $\ptass > 0.3$ \GeVc, $0.3 < \ptass < 1$~\GeVc, $1 <\ptass< 2$~\GeVc, and $2 <\ptass < 3$~\GeVc, significantly extending both transverse momentum coverages with respect to the previous measurements reported in~\cite{ALICE:2016clc}.

The two-dimensional correlation functions are affected by the limited detector acceptance and reconstruction efficiency of the associated tracks (A$^{\rm assoc}\times\epsilon^{\rm assoc}$), as well as the variation of those values for prompt D mesons (A$^{\rm trig}\times\epsilon^{\rm trig}$) inside a given $\ptD$ interval. In order to correct for these effects, a weight equal to 1/(A$^{\rm assoc}\times\epsilon^{\rm assoc}) \times 1$/(A$^{\rm trig}\times\epsilon^{\rm trig}$) was assigned to each correlation pair, as described in detail in~\cite{ALICE:2016clc}.
A weight of 1/(A$^{\rm trig}\times\epsilon^{\rm trig}$) was applied also to the entries in the D-meson invariant mass distributions, used for the evaluation of the amount of signal $S_{\rm{peak}}$ and background $B_{\rm{peak}}$ triggers in the peak region.

The two-dimensional correlation function $C(\Dphi, \Delta\eta)_{\rm{peak}}$ also includes correlation pairs obtained by considering D-meson candidates from combinatorial background as trigger particles. This contribution was subtracted by evaluating the per-trigger correlation function obtained selecting D mesons with an invariant mass in the sidebands, $1/B_{\rm{sidebands}}\times C(\Dphi,\Delta\eta)_{\rm{sidebands}}$, and multiplying it by $B_{\rm{peak}}$. The term $B_{\rm{sidebands}}$ is the amount of background candidates in the sideband region, i.e. 4$\sigma < |M - \mu| < 8\sigma$ (5$\sigma < M - \mu < 10\sigma$, for $\Dstar$ mesons) of the invariant mass distributions weighted by the inverse of the prompt D-meson reconstruction efficiency.

\begin{sloppypar}
The event-mixing technique was used to correct the correlation functions $C(\Dphi,\Delta\eta)_{\rm{peak}}$ and $C(\Dphi,\Delta\eta)_{\rm{sidebands}}$ for the limited detector acceptance and its spatial inhomogeneities. The peak and sideband region event-mixing functions $\mathrm{ME}(\Dphi, \Delta\eta)_{\rm{peak}}$ and $\mathrm{ME}(\Dphi, \Delta\eta)_{\rm{sidebands}}$ were evaluated as explained in~\cite{ALICE:2016clc}. The inverse of these functions was used to weight the functions $C(\Dphi,\Delta\eta)_{\rm{peak}}$ and $C(\Dphi,\Delta\eta)_{\rm{sidebands}}$, respectively.
\end{sloppypar}

The per-trigger angular-correlation function was obtained by subtracting the sideband-region correlation function from the peak-region one, as follows:
\begin{equation}
\label{eq:Corr}
\centering
    \Tilde{C}_{\rm{inclusive}}(\Dphi, \Delta\eta) = \frac{p_\mathrm{prim}(\Dphi)}{S_\mathrm{peak}}\left( \frac{C(\Dphi, \Delta\eta)}{\mathrm{ME}(\Dphi, \Delta\eta)} \bigg|_\mathrm{peak} - \frac{B_\mathrm{peak}}{B_\mathrm{sidebands}}\frac{C(\Dphi, \Delta\eta)}{\mathrm{ME}(\Dphi, \Delta\eta)} \bigg|_\mathrm{sidebands}\right). 
\end{equation}
The division by $S_\mathrm{peak}$ provides the normalisation to the number of D mesons. In our notation per-trigger quantities are specified by the $\Tilde{C}$ symbol.
In Eq.~\ref{eq:Corr}, $p_{\rm{prim}}(\Dphi)$ is a correction for the residual contamination of non-primary associated particles not rejected by the track selection (purity correction).
This was evaluated with Monte Carlo simulations based on PYTHIA6 (Perugia-2011 tune) by quantifying the fraction of primary particles, among all the tracks satisfying the selection criteria.
The correction was applied differentially in $\Dphi$, since from Monte Carlo studies it was verified that this contamination shows a $\Dphi$ modulation, typically of about 1--2\%. The largest value of the contamination was found in the near-side region, for the lowest $\pt$ range of the associated tracks, where $p_{\rm{prim}}(\Dphi)$ approaches 95\%.

Statistical fluctuations prevented a ($\Dphi$,$\Delta\eta$)-double-differential study of the correlation peak properties. Therefore, the per-trigger azimuthal-correlation function $\Tilde{C}_{\rm{inclusive}}(\Dphi)$ was obtained by integrating $\Tilde{C}_{\rm{inclusive}}(\Dphi,\Delta\eta)$ in the range $|\Delta\eta| < 1$.

A fraction of reconstructed D mesons originates from the decay of beauty hadrons (feed-down D mesons).
It was verified with Monte Carlo simulations that azimuthal correlations of prompt and feed-down D mesons with charged particles show different functions. This is a result of the different fragmentation of beauty and charm quarks, as well as of the additional presence of beauty-hadron decay particles in the correlation function of feed-down D-meson triggers.
The contribution of feed-down D-meson triggers to the measured angular-correlation function was subtracted using templates of the azimuthal-correlation function of feed-down D mesons with charged particles, obtained with Monte Carlo simulations at generator level (i.e.~without detector effects and particle selection), as detailed in~\cite{ALICE:2016clc}.

Before performing this subtraction, $\Tilde{C}_{\rm{inclusive}}(\Dphi)$ has to be corrected for a bias which distorts the shape of the near-side region of the feed-down contribution, induced by the D-meson topological selection.
For feed-down D-meson triggers, indeed, the selection criteria are more likely to be satisfied by decay topologies with small angular opening between the trigger D meson and the other products of the beauty-hadron decay. This induces an enhancement of correlation pairs from feed-down D-meson triggers at $\Dphi \approx 0$ and a depletion at larger $\Dphi$ values.
This bias was accounted for as a systematic uncertainty in~\cite{ALICE:2016clc}.
In this paper, instead, a $\Dphi$ dependent correction factor ($c_{\rm{FD-bias}}(\Dphi)$) was determined by comparing Monte-Carlo templates of feed-down D mesons and associated particles at generator level and after performing the event reconstruction and particle selection as on data. This correction factor ranges between 0.6 at $\Dphi \approx 0$ and 1.3 at $\Dphi \approx \pi/4$, decreasing then to 1, and was applied to the feed-down contribution to $\Tilde{C}_{\rm{inclusive}}(\Dphi)$ as follows, to restore this contribution to an unbiased value:
\begin{equation}
\label{Corr_Bias}
    \Tilde{C}^{\rm{corr}}_{\rm{inclusive}}(\Dphi) = \Tilde{C}_{\rm{inclusive}}(\Dphi)\left[\frac{A^{\rm{prompt}}_{\rm NS}(\Dphi)}{A^{\rm{total}}_{\rm NS}(\Dphi)}\times f_{\rm{prompt}}+\frac{A^{\rm{feed-down}}_{\rm NS}(\Dphi)}{A^{\rm{total}}_{\rm NS}(\Dphi)}\times (1-f_{\rm{prompt}})\times c_{\rm{FD-bias}}(\Dphi)\right].
\end{equation}
In Eq.~\ref{Corr_Bias}, $A^{\rm{prompt}}_{\rm NS}(\Dphi)$ ($A^{\rm{feed-down}}_{\rm NS}(\Dphi)$) is the value of the per-trigger correlation function of prompt (feed-down) D-mesons with associated particles, and the term $A^{\rm{total}}_{\rm NS}(\Dphi)$ is the value of the per-trigger correlation function considering both prompt and feed-down components.
The terms $A^{\rm{prompt}}_{\rm NS}(\Dphi)$ and $A^{\rm{feed-down}}_{\rm NS}(\Dphi)$ were evaluated from an analysis on reconstructed Monte Carlo events, where the reconstruction was performed as on data.
The fraction of prompt D mesons in the raw yields, $f_{\rm{prompt}}$, was evaluated as detailed in~\cite{Acharya:2017jgo}. It typically decreases from 95\% to 90\% with increasing $\ptD$ in the studied transverse-momentum intervals, independently of the collision centrality.
The maximum effect of the correction when applied on $\Tilde{C}_{\rm{inclusive}}(\Dphi)$ is about 5\%, at $\Dphi\approx$ 0, for the lowest D-meson $\pt$ range and the highest $\ptass$ interval. The correction becomes negligible for $\ptD > 8$~\GeVc.
After performing this correction, the feed-down contamination was subtracted as described above.

As a result, the fully-corrected, per-trigger azimuthal-correlation function of prompt D mesons with associated particles was obtained, denoted as $1/N_{\rm D} \times {\rm d}N^{\rm assoc}/{\rm d}\Dphi$ from Fig.~\ref{fig:FitOutcome} onwards.

\subsection{Average and fit to the correlation functions}
The correlation functions obtained from $\Dzero$, $\Dplus$, and $\Dstar$ mesons were averaged using, as weights, the inverse of the quadratic sum of the statistical and D-meson uncorrelated systematic uncertainties, discussed in Sec.~\ref{sec:Systematics}, since the three functions were found to be consistent within uncertainties.
Since the correlation functions are symmetric around $\Dphi$ = 0 and $\Dphi$ = $\pi$, they were reflected in the range ${0 < \Dphi < \pi}$ to reduce statistical fluctuations.
In order to quantify the properties of the average D-meson azimuthal-correlation function, it was fitted with the following function:
\begin{equation}
\label{equ:Fit}
    f(\Dphi) = b + \frac{Y_{\rm NS}\times\beta}{2\alpha \Gamma(1/\beta)}\times e^{-\left(\frac{\Dphi}{\alpha}\right)^\beta} + \frac{Y_{\rm AS}}{\sqrt{2\pi}\sigma_{\rm AS}}\times e^{-\frac{(\Dphi - \pi)^2}{2\sigma^2_{\rm AS}}}.
\end{equation}

\begin{figure}[tb]
    \begin{center}
    \includegraphics[width = 0.48\textwidth]{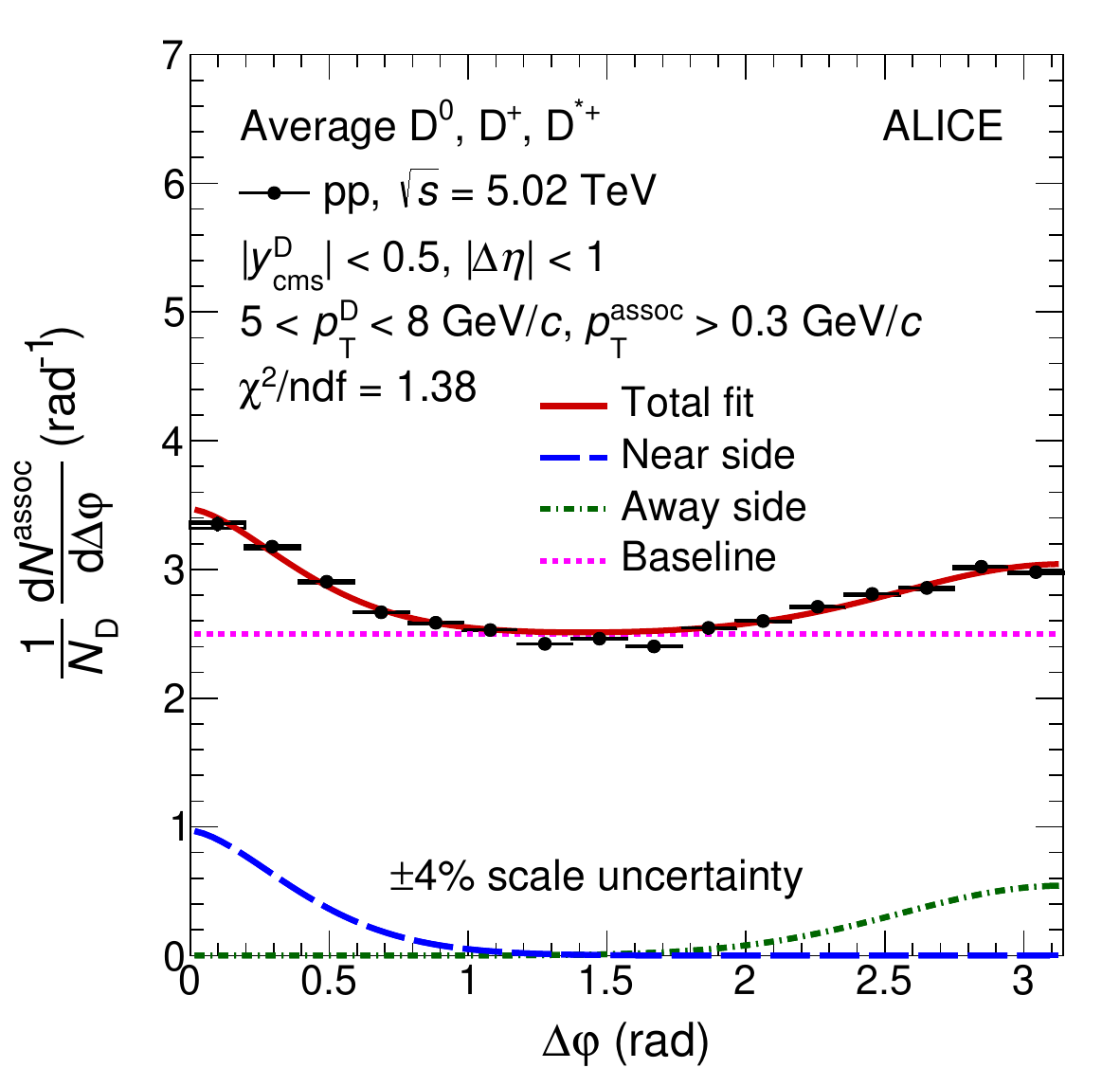}
    \includegraphics[width = 0.48\textwidth]{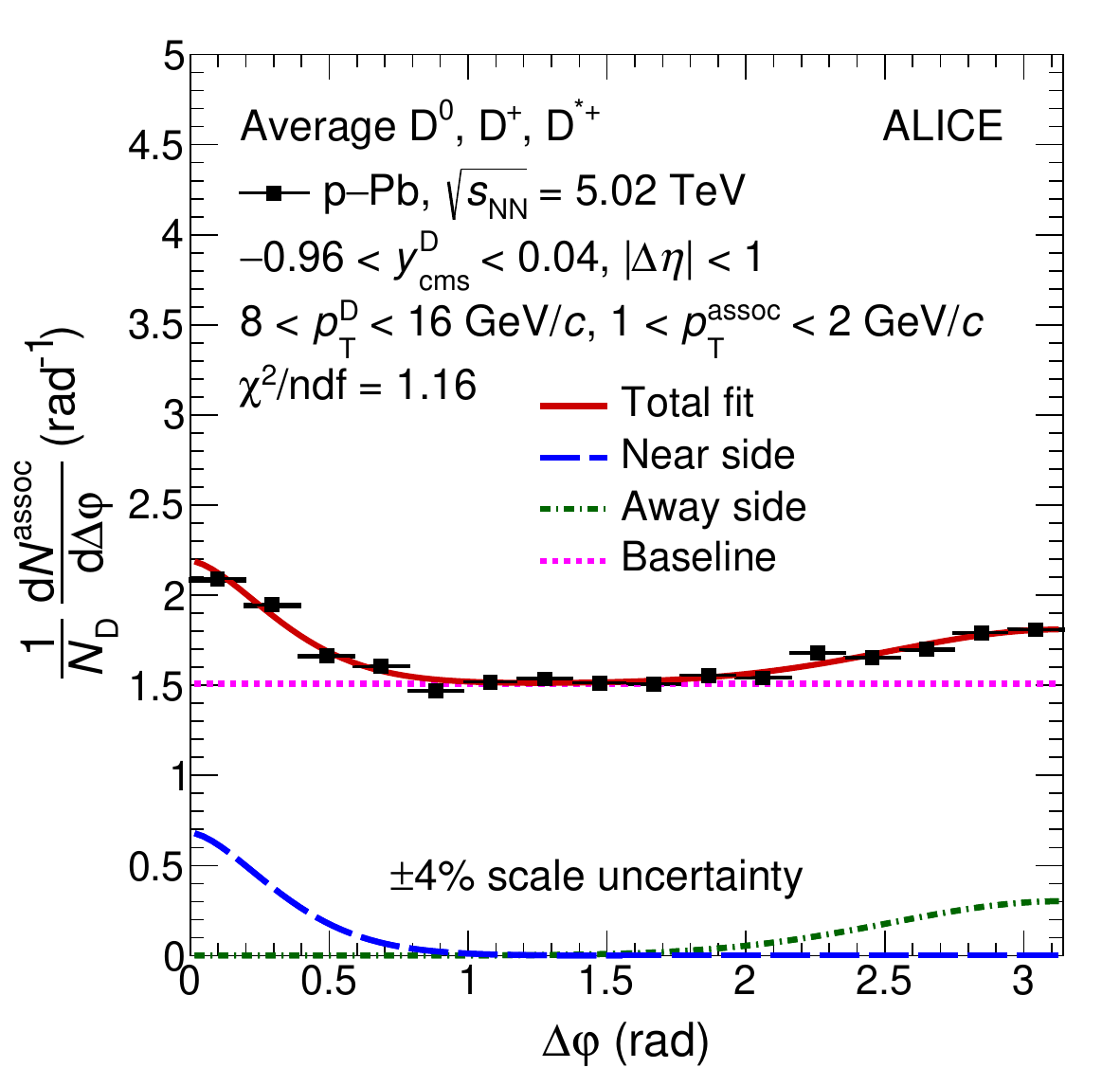}
    \end{center}
    \caption{Examples of the fit to the D-meson average azimuthal-correlation function, for $5 < \ptD < 8$~\GeVc, $\ptass > 0.3$~\GeVc in pp collisions (left), and for $8 < \ptD < 16$~\GeVc, $1 < \ptass < 2$~\GeVc in \pPb collisions (right). The statistical uncertainties are shown as vertical error bars. The fit function described in Eq.~\ref{equ:Fit} is shown as a red solid curve. Its different terms are shown separately: near-side generalised Gaussian function, away-side Gaussian function, and baseline constant term. The scale uncertainty (see Sec.~\ref{sec:Systematics}) is also reported for completeness.}
    \label{fig:FitOutcome}
\end{figure}

The fit function is composed of a constant term $b$ describing the flat contribution below the correlation peaks, a generalised Gaussian term describing the NS peak, and a Gaussian reproducing the AS peak. In the generalised Gaussian, the term $\alpha$ is related to the variance of the function, hence to its width, while the term $\beta$ drives the shape of the peak (the Gaussian function is obtained for $\beta = 2$). The function in Eq.~\ref{equ:Fit} is a generalisation of that adopted in~\cite{ALICE:2016clc}, where a Gaussian function was used for the near-side, corresponding to the case $\beta$ = 2. The new parametrisation allowed to improve the $\chi^2$/ndf value in all the kinematic ranges studied, especially in the high $\pt$ ranges of both the D mesons and associated particles, where the standard Gaussian fit systematically underestimates the near-side peak yields (widths) up to 10\% (20\%) with respect to the generalised Gaussian.
In both collision systems, the $\beta$ parameter decreases monotonically from $\approx 2.2$ at low $\ptD$ and $\ptass$ to $\approx 1$ in the highest $\ptD$ and $\ptass$ intervals.
By symmetry considerations, the means of the Gaussian functions were fixed to $\Dphi$ = 0 and $\Dphi$ = $\pi$.
Figure~\ref{fig:FitOutcome} shows examples of fits to the azimuthal-correlation functions of D mesons with associated particles, for $5 < \ptD < 8$~\GeVc with $\ptass > 0.3$~\GeVc in \pp collisions and for $8 < \ptD < 16$~\GeVc with $1 < \ptass < 2$~\GeVc in \pPb collisions.

\begin{sloppypar}
The integrals of the functions describing the near- and away-side peaks, $Y_{\rm NS}$ and $Y_{\rm AS}$, correspond to the associated-particle yields (i.e. the average number of associated particles contained in the peak), while the widths of the correlation peaks are described by the square root of the variance of their fitting terms, $\alpha\sqrt{\Gamma(3/\beta)/\Gamma(1/\beta)}$ and $\sigma_{\rm AS}$, for the near- and away-side, respectively. The baseline $b$ represents the physical minimum of the $\Dphi$ function, and depends on the average charged-particle multiplicity.
\end{sloppypar}

To reduce the effect of statistical fluctuations on the estimate of the associated yields, $b$ was fixed to the weighted average of the points in the transverse region, defined as $\pi/4 < |\Dphi| < \pi/2$, using the inverse of the point squared statistical uncertainties as weights.

\section{Systematic uncertainties}
\label{sec:Systematics}

The systematic uncertainty induced on the correlation function from the evaluation of $S_{\rm{peak}}$ and $B_{\rm{peak}}$, obtained by fitting the D-meson invariant-mass distribution, was evaluated by varying the fit procedure. In particular, the fit was repeated modelling the background distribution with a linear function and a second-order polynomial function instead of an exponential function (for $\Dzero$ and $\Dplus$ mesons only), varying the fit range, fixing the mean of the Gaussian term describing the mass peak to the world-average D-meson mass~\cite{Tanabashi:2018oca}, or fixing the Gaussian width to the value obtained from Monte Carlo studies. A systematic uncertainty ranging from 1 to 3\% (1 to 2\%), depending on the $\ptD$, was estimated from the corresponding variation of the azimuthal-correlation function for \pp (\pPb) collisions. No dependence on $\Delta\varphi$ was observed and the same uncertainty was estimated for all D-meson species.

An uncertainty ranging from 1 to 3\%, depending on $\ptD$ and on the D-meson species, was assigned in both \pp and \pPb collisions for the possible dependence of the shape of background correlation function on the invariant-mass value of the trigger D meson. This source of uncertainty was determined by evaluating $\Tilde{C}(\Delta\varphi, \Delta\eta)_{\rm{sidebands}}$, defining a different invariant-mass sideband range, and also considering, for $\Dzero$ and $\Dplus$ mesons, only the left or only the right sideband for the evaluation of $\Tilde{C}(\Delta\varphi, \Delta\eta)_{\rm{sidebands}}$. No significant dependence on $\Delta\varphi$ was obtained for this uncertainty.

A systematic effect originating from the correction of the D-meson reconstruction efficiency, due to possible differences of the topological variable distributions between Monte Carlo and data, was evaluated by repeating the analysis applying tighter and looser topological selections on the D-meson candidates, with a corresponding variation of the D-meson reconstruction efficiencies larger than $\pm 25\%$. An uncertainty up to 2.5\% (2\%), increasing for smaller $\ptD$ values, was assigned in \pp (\pPb) collisions. No significant dependence on $\Delta\varphi$ was observed. The same uncertainty was estimated for the three D-meson species.

The systematic uncertainty originating from the evaluation of the associated track reconstruction efficiency was estimated by varying the quality selection criteria applied to the reconstructed tracks, removing the request of at least two associated clusters in the ITS, or requiring a hit on at least one of the two SPD layers, or varying the request on the number of space points reconstructed in the TPC. An uncertainty up to 4.5\% (3\%), was assessed for \pp (\pPb) collisions. No significant trend in $\Delta\varphi$ was observed.

The uncertainty on the evaluation of the residual contamination from secondary tracks was determined by repeating the analysis varying the selection on the DCA in the $xy$ plane from 0.1 to 1 cm, and re-evaluating the purity of associated primary particles for each  variation. This resulted in a 2\% (3\%) maximum systematic uncertainty on the azimuthal-correlation functions in \pp (\pPb) collisions, decreasing with increasing $\ptass$ and with negligible $\Dphi$ dependence.

The uncertainty on the subtraction of the beauty feed-down contribution was quantified by generating the templates of feed-down azimuthal-correlation functions with different event generators (PYTHIA6 with the Perugia-2010 tune, PYTHIA8 with the 4C tune) and by varying the value of $f_{\rm{prompt}}$ within its uncertainty band, as described in details in~\cite{Acharya:2019mgn}. The resulting uncertainty was found to be dependent on $\Delta\varphi$, with a maximum value of 5\% (3\%) in \pp (\pPb) collisions, and was applied point-by-point on the correlation functions.

As discussed in Sec.~\ref{sec:Analysis}, Monte Carlo studies revealed the presence of a bias on the near-side region of the correlation function for feed-down D-mesons triggers, induced by the topological selections applied to the D mesons. The correction applied to remove this bias relies on a proper description of the azimuthal-correlation functions of prompt and feed-down D-meson triggers by the Monte Carlo simulations. A $\Dphi$-dependent, symmetric systematic uncertainty of $\pm\delta\Tilde{C}(\Delta\varphi)/\sqrt{12}$
was introduced to account for under- or overestimation of the correction, where $\delta\Tilde{C}(\Delta\varphi)$ is the point-by-point shift of the correlation function induced by the correction. The largest value of the uncertainty was 2\%, at $\Dphi \approx 0$, for both \pp and \pPb collisions.

In Tab.~\ref{Syst_Table}, the minimum and maximum values of the systematic uncertainties affecting the azimuthal-correlation functions, depending on the kinematic range, are listed for both collision systems. Only the uncertainties deriving from the feed-down subtraction and from the correction on the bias of feed-down D-meson correlations are $\Dphi$ dependent. All the other contributions define a $\Dphi$-independent systematic uncertainty, which acts as a scale uncertainty for the correlation function. In both pp and $\pPb$ collisions the total scale uncertainty ranges from $\pm$4\% to $\pm$5\%.

\begin{table}[h]
\centering
\begin{tabular}{l|c|c}
\textbf{System} & \textbf{pp} & \textbf{\pPb}\\
D-meson species & $\Dzero$, $\Dstar$, $\Dplus$ & $\Dzero$, $\Dstar$, $\Dplus$\\
\hline
Signal, background normalisation            & $\pm$1--3\%        & $\pm$1--2\% \\
Background $\Dphi$ function                 & $\pm$1--3\%        & $\pm$1--3\%  \\
Associated-track reconstruction efficiency  & $\pm$2.5--4.5\%    & $\pm$3\% \\
Primary-particle purity                     & $\pm$1--2\%        & $\pm$1.5--3\% \\
D-meson efficiency                          & $\pm$1--2.5\%      & $\pm$1--2\% \\
Feed-down subtraction                       & up to 5\%, $\Dphi$-dependent & up to 3\%, $\Dphi$-dependent \\
Bias on topological selection               & up to 2\%, $\Dphi$-dependent & up to 2\%, $\Dphi$-dependent\\
\hline
\end{tabular}
\caption{List of systematic uncertainties for the azimuthal-correlation functions in pp and in $\pPb$ collisions. If not specified, the uncertainty does not depend on $\Dphi$.}
\label{Syst_Table}
\end{table}

The systematic uncertainties on the near- and away-side peak yields and widths, and on the baseline height, obtained from the fits to the azimuthal-correlation functions, were evaluated as follows.
The main source of uncertainty arises from the definition of the $\Delta\varphi$ transverse region used to determine the baseline height (term $b$ of Eq.~\ref{equ:Fit}).
The impact on the physical observables induced by the baseline value was estimated by considering different $\Delta\varphi$ ranges for determining the baseline position and performing the fits again using Eq.~\ref{equ:Fit}. Moreover, the fits were repeated by moving the points of the correlation functions upwards and downwards using the corresponding value of the $\Delta\varphi$-dependent systematic uncertainty.
The total systematic uncertainty was calculated by summing in quadrature the aforementioned contributions. For the associated yields and for the baseline, whose values depend on the normalisation of the correlation function, also the $\Delta\varphi$-independent systematic uncertainties affecting the correlation function (i.e.~the first five contributions listed in Table~\ref{Syst_Table}), which act as a scale factor, were summed in quadrature.

In $\pPb$ collisions, the presence of long-range correlations among the particles produced in the collision can have an impact on the values of the quantities extracted from the fits, in particular for the analysis as a function of centrality. This effect was studied by fitting the functions with a $v_{2\Delta}$-like modulation~\cite{Acharya:2018dxy}, in place of a flat baseline. The $v_2$ values adopted for D mesons, ranging up to 8\% for the lowest $\ptD$ range in 0--20\% central events, were estimated employing the available results for heavy-flavour particle $v_2$ in \pPb collisions from CMS~\cite{Sirunyan:2018toe}, ALICE~\cite{Acharya:2018dxy}, and ATLAS~\cite{ATLAS-CONF-2017-073,ATLAS-CONF-2017-006}, while those for associated particles were estimated based on di-hadron correlation measurements by ALICE~\cite{Abelev:2012ola}.
For the centrality-integrated analysis and for the case when $\ptass > 0.3$~\GeVc, considering a $v_{2\Delta}$-like modulation reduced the near-side peak yields by about 16\% (5\%) for $3 < \ptD < 5$~\GeVc ($5 < \ptD < 8$~\GeVc) and the away-side peak yields by about 20\% (3\%) for $3 < \ptD < 5$~\GeVc ($5 < \ptD < 8$~\GeVc). A smaller variation was observed for the peak widths and for the baseline value.
For the analysis as a function of the event centrality, the largest effect was obtained for the 0--20\% centrality class, where for $\ptass > 0.3$~\GeVc a decrease of 27\%, 17\%, and 5\% was found for $3 < \ptD < 5$~\GeVc, $5 < \ptD < 8$~\GeVc, and ${8 < \ptD < 16}$~\GeVc, respectively. Smaller variations were found for the near-side peak width and the baseline.
This systematic uncertainty was summed in quadrature with the others to obtain the total uncertainty.

\section{Results}
\label{sec:Results}

\subsection{Comparison of results in \pp and \pPb collisions}
\label{subsec:ppvspPb}

\begin{figure}[tb]
    \begin{center}
    \includegraphics[width = 0.99\textwidth]{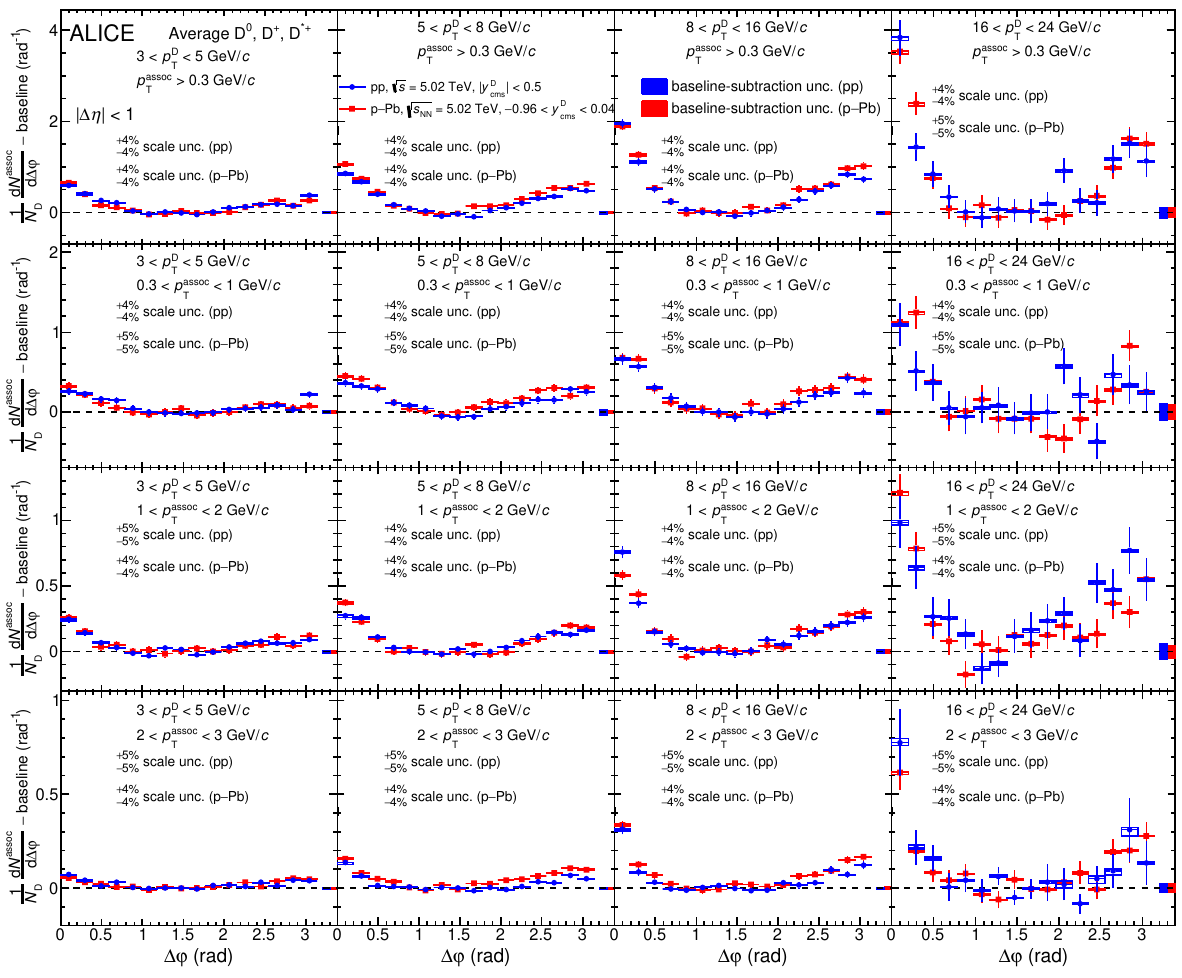}
    \end{center}
    \caption{Average of the azimuthal-correlation functions of $\Dzero$, $\Dplus$, and $\Dstar$ mesons with associated particles, after the subtraction of the baseline, in \pp collisions at ${\s = 5.02}$ \TeV and \pPb collisions at $\snn = 5.02$~\TeV, for $3 < \ptD < 5$~\GeVc, $5 < \ptD < 8$~\GeVc, $8 < \ptD < 16$~\GeVc, and $16 < \ptD < 24$~\GeVc  (from left to right) and $\ptass > 0.3$~\GeVc, $0.3 < \ptass < 1$~\GeVc, $1 < \ptass < 2$~\GeVc, and $2 < \ptass < 3$~\GeVc (from top to bottom). Statistical and $\Dphi$-dependent systematic uncertainties are shown as vertical error bars and boxes, respectively, $\Dphi$-independent uncertainties are written as text. The uncertainties from the subtraction of the baseline are displayed as boxes at $\Dphi > \pi$.}
    \label{fig:ppvspPb_dPhi}
\end{figure}

The averaged azimuthal-correlation functions of the $\Dzero$, $\Dplus$, and $\Dstar$ mesons with associated particles in \pp and \pPb collision systems are compared, after baseline subtraction, in Fig.~\ref{fig:ppvspPb_dPhi}, for four D-meson transverse momentum ranges, ${3 < \ptD < 5}$~\GeVc, ${5 < \ptD < 8}$~\GeVc, ${8 < \ptD < 16}$~\GeVc, and ${16 < \ptD < 24}$~\GeVc. The functions are presented for $\ptass > 0.3$~\GeVc as well as for three sub-ranges, $0.3 < \ptass < 1$~\GeVc, $1 < \ptass < 2$~\GeVc, and $2 < \ptass < 3$~\GeVc.
The qualitative shape of the correlation function and the evolution of the near- and away-side peaks with trigger and associated particle $\pt$ are consistent within uncertainties in the two collision systems. In particular, an increase of the height of the near-side correlation peak is observed for increasing values of the D-meson \pt. This reflects the production of a higher number of particles in the jet accompanying the fragmenting charm quark, when the energy of the latter increases. A similar, though milder, effect can be observed also for the away-side peak.

\begin{figure}[tb]
    \begin{center}
    \includegraphics[width = 0.99\textwidth]{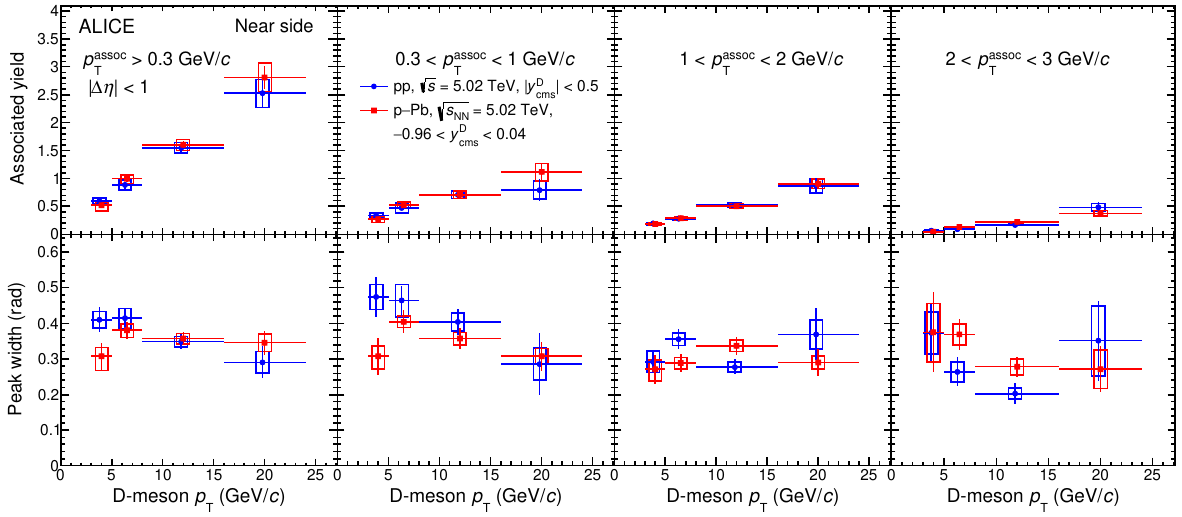}
    \end{center}
    \caption{Near-side peak associated yields (top row) and widths (bottom row) in \pp collisions at ${\s = 5.02}$ \TeV and \pPb collisions at ${\snn = 5.02}$ \TeV, as a function of the D-meson $\pt$, for ${\ptass > 0.3}$~\GeVc, ${0.3 < \ptass < 1}$~\GeVc, ${1 < \ptass < 2}$~\GeVc, and ${2 < \ptass < 3}$~\GeVc (from left to right). Statistical and systematic uncertainties are shown as vertical error bars and boxes, respectively. The points and error boxes for pp collisions are shifted by $\Delta\pt = -0.2$~\GeVc.} 
    \label{fig:ppvspPb_NS}
\end{figure}

A more quantitative comparison of the near- and away-side peak features and $\pt$ evolution in the two collision systems can be obtained by fitting the azimuthal-correlation functions and evaluating the peak yields and widths, as it was explained in Sec.~\ref{sec:Analysis}.
Figure~\ref{fig:ppvspPb_NS} compares these observables for the near-side correlation peaks in \pp and \pPb collisions, as a function of the D-meson $\pt$, for $\ptass > 0.3$~\GeVc and in three $\ptass$ sub-ranges. For both yields and widths, the values measured in the two collision systems are in agreement. The increase of associated particle production inside the near-side peak with $\ptD$, qualitatively observed in Fig.~\ref{fig:ppvspPb_dPhi}, is present for all the associated particle $\pt$ intervals, and is similar in the two collision systems.
A tendency for a narrowing of the near-side peak with increasing $\ptD$ is also observed in most of the $\pt$ ranges, though a flat behaviour cannot be excluded with the current uncertainties.

\begin{figure}[tb]
    \begin{center}
    \includegraphics[width = 0.99\textwidth]{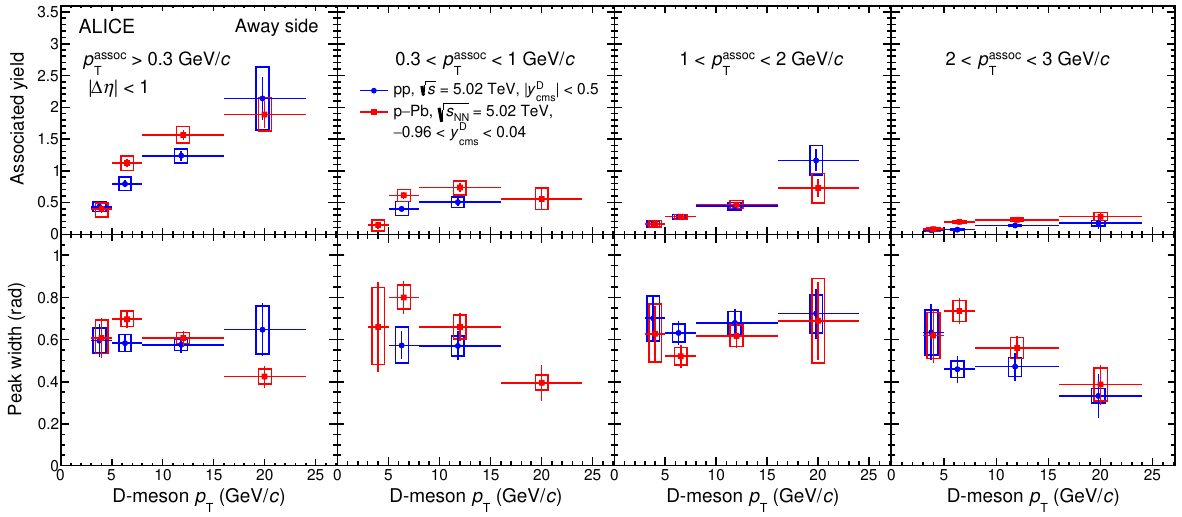}
    \end{center}
    \caption{Away-side peak associated yields (top row) and widths (bottom row) in \pp collisions at ${\s = 5.02}$ \TeV and \pPb collisions at ${\snn = 5.02}$ \TeV, as a function of the D-meson $\pt$, for ${\ptass > 0.3}$~\GeVc, ${0.3 < \ptass < 1}$~\GeVc, ${1 < \ptass < 2}$~\GeVc, and ${2 < \ptass < 3}$~\GeVc (from left to right). Statistical and systematic uncertainties are shown as vertical error bars and boxes, respectively. The points and error boxes for pp collisions are shifted by $\Delta\pt = -0.2$~\GeVc.}
    \label{fig:ppvspPb_AS}
\end{figure}

The away-side peak yields and widths measured in \pp and \pPb collisions are compared in Fig.~\ref{fig:ppvspPb_AS} as a function of the D-meson $\pt$, with the common associated-particle $\pt$ ranges analysed.
For \pp collisions, specific kinematic regions where the $\chi^2/$ndf of the fit was much larger than unity, or where the uncertainties on the peak observables were larger than 100\%, were excluded from the results.
As in the near-side analysis, the away-side yields show an increasing trend with $\ptD$, and overall have similar values in the two collision systems. In the intermediate D-meson transverse momentum range, there is a hint for larger yields in \pPb than in \pp, but not a statistically significant one (about 2.2$\sigma$ for the combined range $5 < \ptD < 16$~\GeVc for all $\ptass$ ranges). The away-side peak widths show consistent values in \pp and \pPb collisions in all kinematic ranges.
No significant impact from cold-nuclear-matter effects on the fragmentation and hadronisation of charm quarks appears from the comparison of the results in the two collision systems, within the current precision of the measurements. This result complements the observation, emerged from the measurements reported in Refs.~\cite{Acharya:2019mno,Adam:2015qda}, that cold-nuclear-matter effects have a small impact on the production of charm quarks at midrapidity in \pPb collisions.

\subsection{Results in \pPb collisions as a function of the event centrality}
\begin{figure}[tb]
    \begin{center}
    \includegraphics[width = 0.99\textwidth]{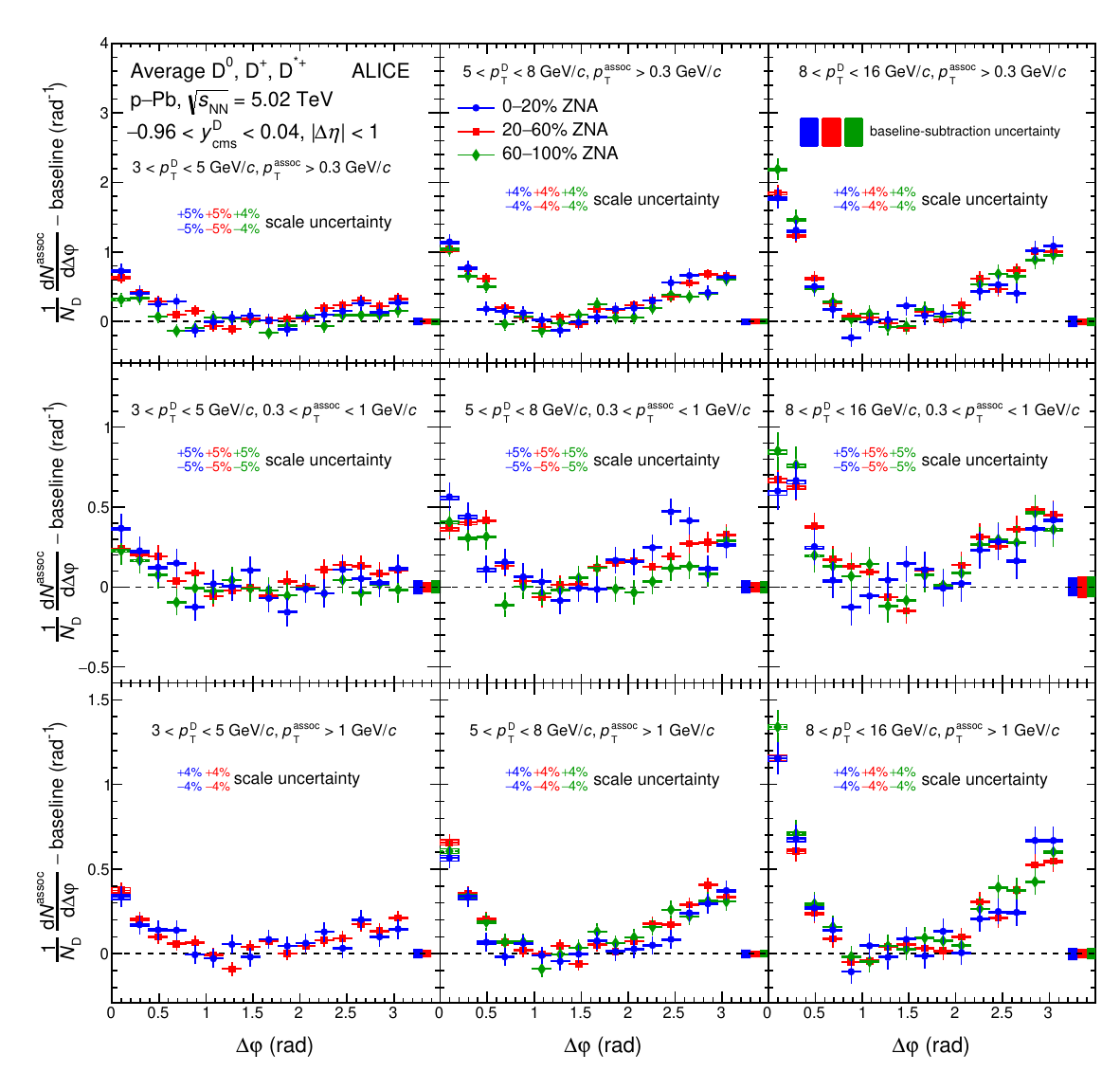}
    \end{center}
    \caption{Average of the azimuthal-correlation functions of $\Dzero$, $\Dplus$, and $\Dstar$ mesons with associated particles, after the subtraction of the baseline, for \pPb collisions in three different centrality classes, 0--20\% (blue circles), 20--60\% (red squares), and 60--100\% (green diamonds). The functions are shown for $3 < \ptD < 5$~\GeVc, $5~<~\ptD~<~8$~\GeVc, and $8 < \ptD < 16$~\GeVc (from left to right) and $\ptass > 0.3$~\GeVc, $0.3 < \ptass < 1$~\GeVc, and $\ptass > 1$~\GeVc (from top to bottom). Statistical and $\Dphi$-dependent systematic uncertainties are shown as vertical error bars and boxes, respectively, while $\Dphi$-independent uncertainties are written as text. The uncertainties from the subtraction of the baseline are displayed as boxes at $\Dphi > \pi$.}
    \label{fig:pPbvsCent_dPhi}
\end{figure}

\begin{sloppypar}
The correlation functions of D mesons with associated particles for \pPb collisions in the 0--20\%, 20--60\%, and 60--100\% centrality classes are compared in Fig.~\ref{fig:pPbvsCent_dPhi}, for nine kinematic ranges with ${3 < \ptD < 16}$~\GeVc and ${\ptass > 0.3}$~\GeVc.
No results are shown for the 60--100\% centrality class, for ${3 < \ptD < 5}$~\GeVc and ${\ptass > 1}$~\GeVc, because of instabilities in the fits to the correlation functions induced by statistical fluctuations.
For the comparison of the correlation peak characteristics, the baseline values were subtracted from the functions, since they strongly depend on the centrality interval. For the $\ptass > 0.3$~\GeVc interval, the baseline values lied in the ranges 7.7--8, 6.2--6.6, and 4--4.2, for the 0--20\%, 20--60\%, and 60--100\% centrality classes, respectively, showing no dependence with the D-meson $\pt$.
The baseline-subtracted correlation functions do not show significant differences among the three centrality intervals studied.
\end{sloppypar}

\begin{figure}
    \begin{center}
    \includegraphics[width = 0.99\textwidth]{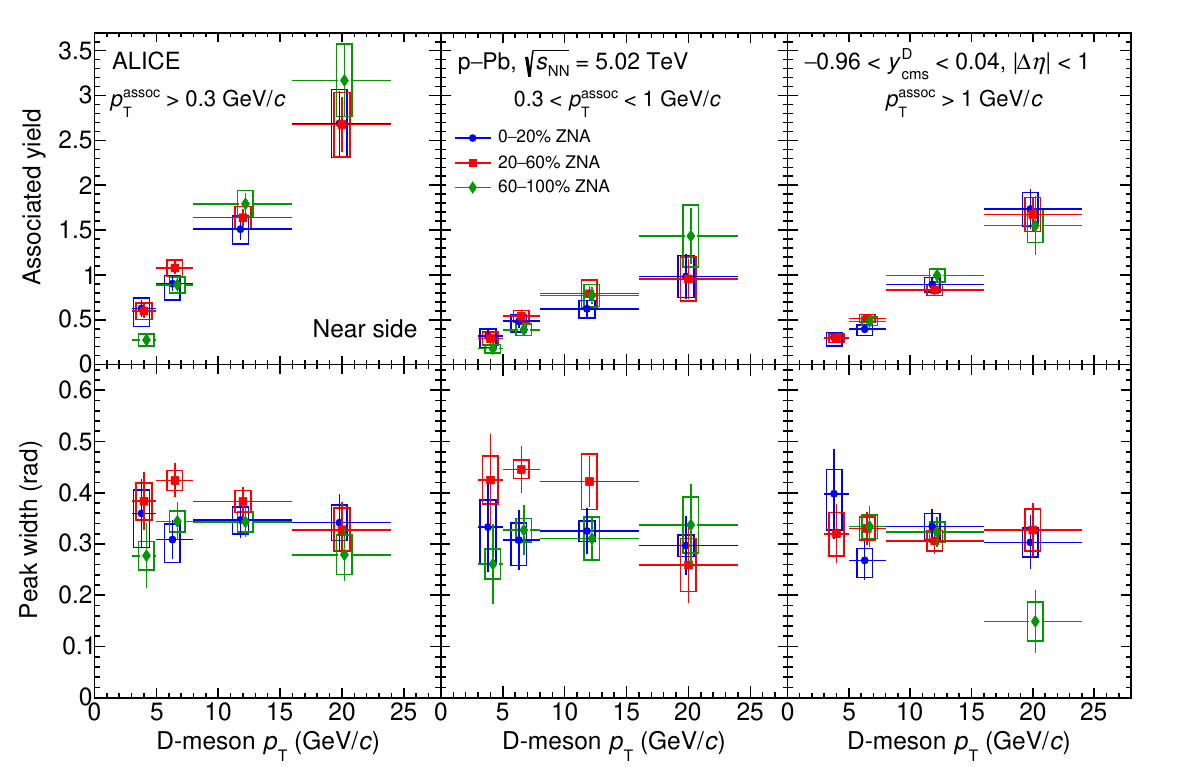}
    \end{center}
    \caption{Comparison of near-side associated peak yields (top row) and widths (bottom row) as a function of the D-meson $\pt$, for \pPb collisions in three different centrality classes, 0--20\% (blue circles), 20--60\% (red squares), and 60--100\% (green diamonds). The results are presented as a function of the D-meson $\pt$, for $\ptass > 0.3$~\GeVc, $0.3 < \ptass < 1$~\GeVc, and $\ptass > 1$~\GeVc (from left to right). Statistical and systematic uncertainties are shown as vertical error bars and boxes, respectively. The points and error boxes for 0--20\% (60--100\%) collisions are shifted by $\Delta\pt = -0.2$ ($+0.2$)~\GeVc}
    \label{fig:pPbvsCent_NS}
\end{figure}

Figure~\ref{fig:pPbvsCent_NS} shows the near-side yields and widths extracted by a fit to the correlation functions, for the three centrality intervals.
A similar increase of the near-side peak yields, as a function of the D-meson $\pt$, is observed for the three centrality ranges, with the absolute values of the yields also being generally in agreement.
The only exception is for the $3 < \ptD < 5$~\GeVc, $\ptass > 0.3$~\GeVc interval, where the yield for the 60--100\% centrality class is lower than for the 0--20\% and 20--60\% centrality classes, with a statistical significance of $1.4\sigma$ and $2.1\sigma$, respectively. This effect could be due to statistical fluctuations of the correlation function data points (see Fig.~\ref{fig:pPbvsCent_dPhi}).
The near-side peaks also have consistent widths among the three centrality ranges, for all the kinematic ranges studied.
No centrality dependence on the correlation peaks, which could have possibly been induced by nuclear-matter effects, is observed within the experimental uncertainties. The limited precision of the results does not provide a further validation of the subtraction technique of the jet-induced correlation peaks, commonly used in analyses searching for positive elliptic flow via two-particle correlations.

\subsection{Comparison of ALICE results to predictions from Monte Carlo simulations}
The azimuthal-correlation functions of D mesons with associated particles, as well as the near- and away-side peak yields and widths measured by ALICE in \pp collisions, were compared to expectations from several Monte Carlo event generators.

The PYTHIA event generator~\cite{Sjostrand:2006za,Sjostrand:2007gs} allows for the generation of high-energy collisions of leptons and/or hadrons. It employs 2 $\rightarrow$ 2 QCD matrix elements evaluated perturbatively with leading-order precision, with the next-to-leading order contributions taken into account during the parton showering stage. The parton showering follows a leading-logarithmic $\pt$ ordering, with soft-gluon emission divergences excluded by an additional veto, and the hadronisation is handled with the Lund string-fragmentation model.
Two different versions of PYTHIA, with two different parameter tunes, were used in this paper.
The PYTHIA 6.4.25 version~\cite{Sjostrand:2006za} was employed, incorporating the Perugia 2011 tune~\cite{Skands:2010ak}, which was the first tune considering the data from \pp collisions at $\s = 0.9$~\TeV and $\s = 7$~\TeV at the LHC.
With respect to its predecessor, PYTHIA8~\cite{Sjostrand:2007gs} has an improved handling of the multiple-parton interactions and the colour reconnection processes. In this paper, it was used with the tune 4C~\cite{Corke:2010yf}.

POWHEG~\cite{Nason:2004rx,Frixione:2007vw} is a pQCD generator implementing hard-scattering matrix elements with NLO accuracy, which can be coupled to Monte Carlo generators, like PYTHIA~\cite{Sjostrand:2006za,Sjostrand:2007gs} or HERWIG~\cite{Corcella:2000bw,Bellm:2015jjp}, for the parton showering and hadronisation of the produced partons. In this paper, Monte Carlo simulations were done using the POWHEG-BOX~\cite{Alioli:2010xd} framework coupled to PYTHIA 6.4.25 with the Perugia-2011 tune~\cite{Skands:2010ak}.
A charm-quark mass of $m_{\rm c} = 1.5$~\GeVmass was considered, and the renormalisation and factorisation scales were set to the transverse mass of charm quark, i.e. $\mu_{\rm R} = \mu_{\rm F} = \sqrt{\pt^2 + m_{\rm c}^2}$.
It was verified that simulation results do not change significantly when varying the generator parameters according to the guidelines in~\cite{Cacciari:2012ny}: the variation of the charm-quark mass does not alter the correlation function, while the variation of the renormalisation and factorisation scales produces differences of $\pm10\%$ ($\pm5\%$) for the near-side peak yields (widths) and negligible deviations for the away-side peak yields and widths.
This can be expected, since the per-trigger correlation function of D mesons with associated particles is scarcely sensitive to the absolute rate of production of D mesons, directly influenced by the aforementioned parameters.
An additional set of predictions from POWHEG+PYTHIA was also evaluated (POWHEG LO+PYTHIA6 in the following), by stopping the computation of the hard-scattering matrix elements at leading-order accuracy, before passing the generated partons to PYTHIA for the showering and hadronisation.

The HERWIG 7~\cite{Bahr:2008pv,Bellm:2015jjp} event generator allows one to perform Monte Carlo simulations at NLO accuracy for most of the Standard Model processes, including heavy-quark production. The parton showering is performed with an angular ordering of the fragments, which correctly takes the coherence effects for soft-gluon emissions into account. In addition, the hadronisation is handled via the cluster hadronisation model, differently from the Lund string fragmentation model employed by PYTHIA.

EPOS 3~\cite{Werner:2010aa,Drescher:2000ha} is a Monte Carlo generator which considers flux tube initial conditions for the collision, generated in the Gribov-Regge multiple-scattering framework, and applies a 3+1D viscous hydrodynamical evolution on the dense core of the collision. Individual scatterings, referred to as Pomerons, are identified with parton ladders, each composed of a pQCD hard process, plus initial- and final-state radiations. The hadronisation is then performed with a string fragmentation procedure. Non-linear effects are considered by means of a saturation scale. An evaluation within the EPOS 3 model shows that the energy density reached in pp collisions at the LHC energies is already sufficient for applying such a hydrodynamic evolution~\cite{Werner:2013tya}.
In the following, due to the limited precision of the available predictions, the comparison between EPOS 3 expectations and data results will be restricted to the kinematic interval $3 < \ptD < 16$~\GeVc, and will not include the away-side peak observables.

\begin{figure}[tb]
    \begin{center}
    \includegraphics[width = 0.99\textwidth]{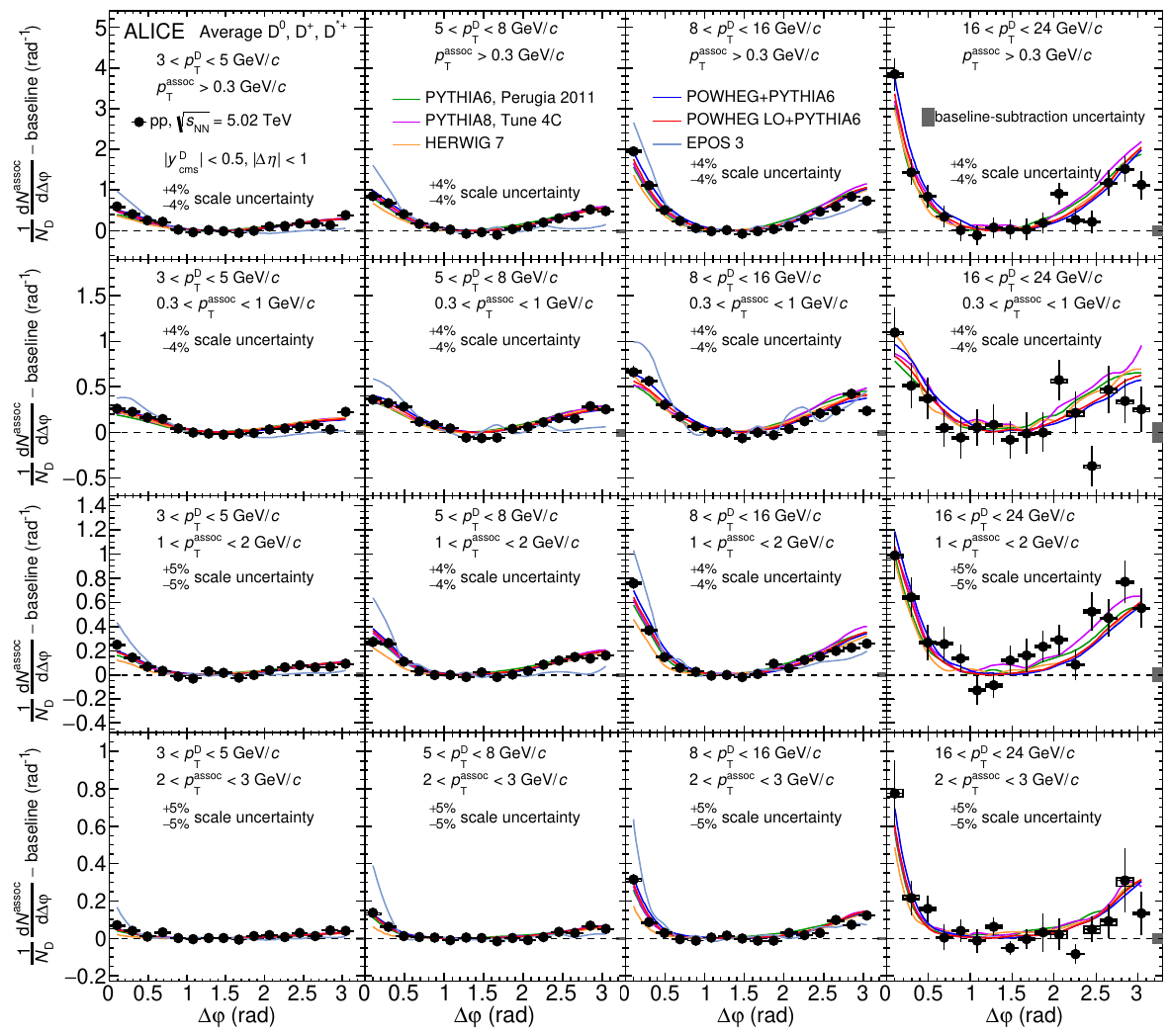}
    \end{center}
    \caption{The average of the azimuthal-correlation functions of $\Dzero$, $\Dplus$, and $\Dstar$ mesons with associated particles, after the subtraction of the baseline, in \pp collisions at $\s = 5.02$~\TeV, compared to predictions from the PYTHIA, POWHEG+PYTHIA6, POWHEG LO+PYTHIA6, HERWIG, and EPOS 3 event generators with various configurations (see text for details). The functions are shown for $3 < \ptD < 5$~\GeVc, $5 < \ptD < 8$~\GeVc, $8 < \ptD < 16$~\GeVc, and $16 < \ptD < 24$~\GeVc  (from left to right) and $\ptass > 0.3$~\GeVc, $0.3 < \ptass < 1$~\GeVc, $1 < \ptass < 2$~\GeVc, and $2 < \ptass < 3$~\GeVc (from top to bottom). Statistical and $\Dphi$-dependent systematic uncertainties are shown as vertical error bars and boxes, respectively, while the $\Dphi$-independent uncertainties are written as text. The uncertainties from the subtraction of the baseline are displayed as boxes at $\Dphi > \pi$.}
    \label{fig:ppvsModels_dPhi}
\end{figure}

In Fig.~\ref{fig:ppvsModels_dPhi} the azimuthal-correlation functions of D mesons with associated particles obtained from the aforementioned event generators are compared to the measurements from this analysis, for all the $\ptD$ and $\ptass$ ranges studied, in \pp collisions at $\s = 5.02$ \TeV, after the baseline subtraction.
For the models, for which the statistical fluctuations are generally negligible, the baseline was estimated as the minimum of the azimuthal-correlation function, and a systematic uncertainty on the fit parameters was assessed by repeating the fits after fixing the baseline as the weighted average of the two lowest points of the correlation function.
Most of the models provide a fair description of the two correlation peaks in the various kinematic ranges studied, though some tensions are visible from this qualitative comparison. In particular, HERWIG underestimates the near-side peak height for $\ptass > 1$~\GeVc, especially for low D-meson transverse momentum, while EPOS 3 tends to overestimate the height of the near-side peak and gives a flatter away-side peak.
In addition, some systematic hierarchies among the models appear throughout the whole $\pt$ ranges analysed, with POWHEG+PYTHIA6 providing the highest near-side peak, and in most of the cases the smallest away-side peak.
The overestimation of the near-side peak yield by EPOS 3 is a relevant feature also for the understanding of the dependence of heavy-flavour production on the charged-particle multiplicity measured in the same rapidity window of the heavy-flavour signals~\cite{Adam:2015ota}. Disentangling the role of jet-biases from effects related to genuine global event properties is fundamental for properly interpreting the measured trends, especially their $\pt$ dependence~\cite{Weber:2018ddv}.

\begin{figure}[tb]
    \begin{center}
    \includegraphics[width = 0.99\textwidth]{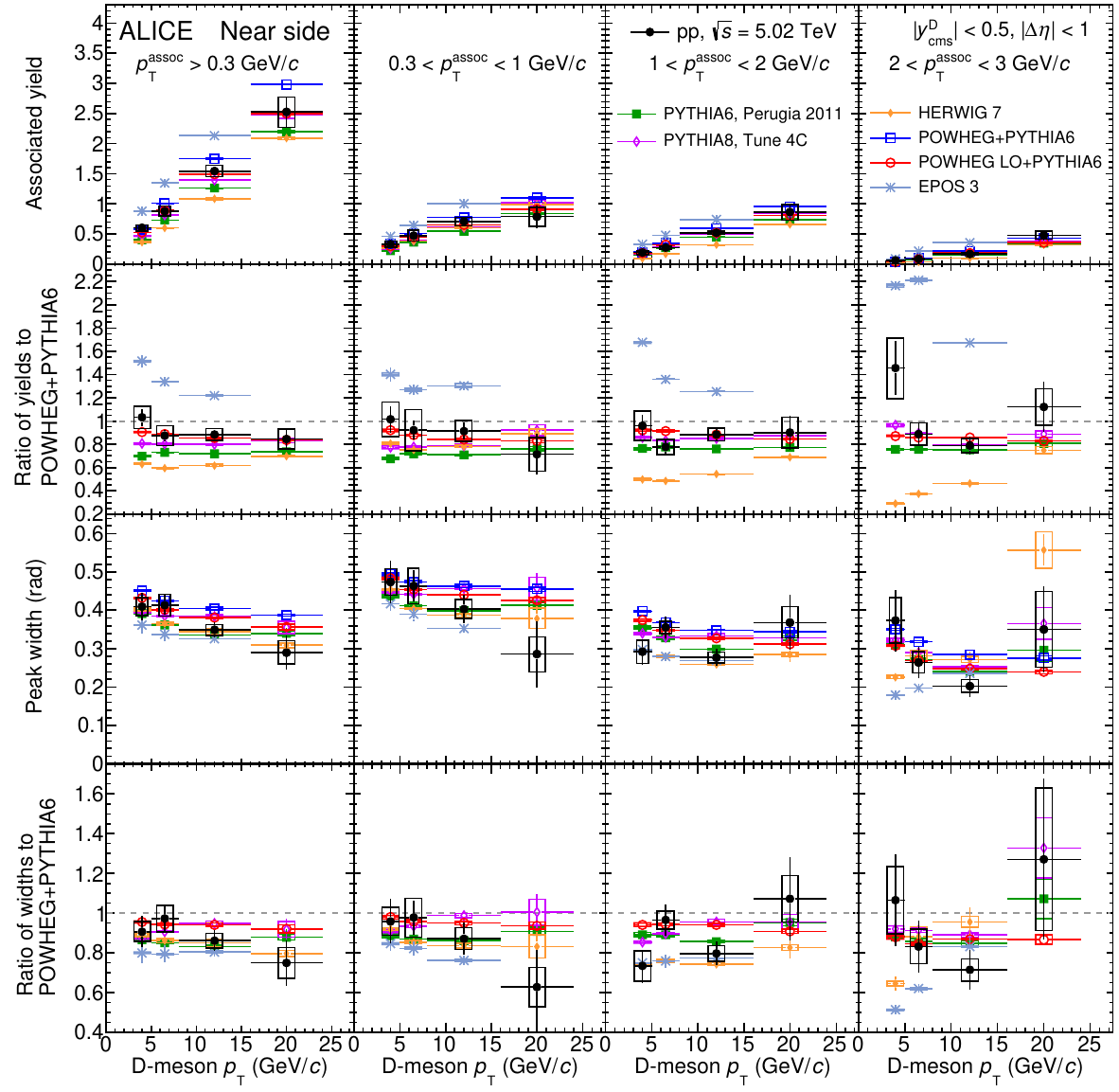}
    \end{center}
    \caption{Measurements of near-side associated peak yields (top row) and widths (third row down) in \pp collisions at $\s = 5.02$~\TeV, compared to predictions by the PYTHIA, POWHEG+PYTHIA6, POWHEG LO+PYTHIA6, HERWIG, and EPOS 3 event generators with various configurations (see text for details). The ratios of yield (width) values with respect to the predictions by POWHEG+PYTHIA6 are shown in the second (fourth) row down. Results are presented as a function of the D-meson $\pt$, for $\ptass > 0.3$~\GeVc, $0.3 < \ptass < 1$~\GeVc, $1 < \ptass < 2$~\GeVc, and $2 < \ptass < 3$~\GeVc (from left to right). Statistical and systematic uncertainties are shown as vertical error bars and boxes, respectively.}
    \label{fig:ppvsModels_NS}
\end{figure}

A more detailed investigation can be performed by quantifying the peak yields and widths extracted from the fit to the correlation functions. In Fig.~\ref{fig:ppvsModels_NS}, the comparison of near-side peak yields and widths from data and simulation is shown, as a function of the D-meson $\pt$, for $\ptass > 0.3$~\GeVc and for the three $\ptass$ sub-ranges analysed, in \pp collisions at $\s = 5.02$~\TeV. In the top row (third row down), the absolute value of the yields (widths) are displayed, while the second (fourth) row down reports the ratios of the yields (widths) to those obtained with POWHEG+PYTHIA6, which reduces the visual impact of the statistical fluctuations of the data points.
As already visible from Fig.~\ref{fig:ppvsModels_dPhi},  EPOS 3 predicts the largest values of the near-side yields, followed by POWHEG+PYTHIA6, while POWHEG LO+PYTHIA6 shows about 10\% lower yields with respect to the version with NLO accuracy. The latter difference could be explained by a different relative contribution of the NLO production mechanisms, in particular the gluon splitting, present already at the level of the hard scattering for POWHEG+PYTHIA6. PYTHIA8 provides near-side yield values comparable to those of POWHEG LO+PYTHIA6, while PYTHIA6 yields are slightly lower. HERWIG expectations for near-side yields are the lowest, except for the $0.3 < \ptass < 1$~\GeVc range, where they are comparable to PYTHIA8 expectations.
POWHEG+PYTHIA6 and POWHEG LO+PYTHIA6 provide the best description of the near-side yields, with data points lying between the two predictions. PYTHIA8 also gives a good description of data, especially for $\ptass > 1$~\GeVc, while PYTHIA6 predictions are generally lower than data, though in agreement within the uncertainties. The HERWIG expectations for near-side yield describe the data well for the $0.3 < \ptass < 1$~\GeVc range, while they severely underpredict the measurements for the other $\ptass$ ranges, especially for the lower intervals of the D-meson $\pt$. In particular, for the integrated range $\ptass > 0.3$~\GeVc, a discrepancy of $3.3\sigma$ ($2.9\sigma$) is found for $3 < \ptD < 5$~\GeVc ($5 < \ptD < 8$~\GeVc), increasing to $3.4\sigma$ ($3.6\sigma$) for the highest associated particle transverse-momentum range $2 < \ptass < 3$~\GeVc.
The EPOS 3 model largely overestimates the near-side associated yields, especially at low D-meson $\pt$. For $\ptass > 0.3$~\GeVc, the discrepancy between data and predictions ranges between $4.0\sigma$ and $5.2\sigma$.
Except for EPOS 3, a similar hierarchy among the models also characterises the near-side widths. POWHEG+PYTHIA6 give the broadest peaks, followed by POWHEG LO+PYTHIA6, with increasing difference between the two model expectations with increasing $\ptass$. PYTHIA8 gives similar widths as POWHEG LO+PYTHIA6, while PYTHIA6 widths are generally lower. The lowest predictions are provided by EPOS 3. HERWIG predictions are consistent with PYTHIA6 for $\ptass < 1$~\GeVc, and are generally lower for $\ptass > 1$~\GeVc.
POWHEG+PYTHIA6 provides systematically larger widths than data, though still being compatible point-by-point. EPOS 3 predictions tend to underestimate the near-side widths, despite being consistent with data point-by-point. All the other models provide values of the near-side width closer to data.

\begin{figure}[tb]
    \begin{center}
    \includegraphics[width = 0.99\textwidth]{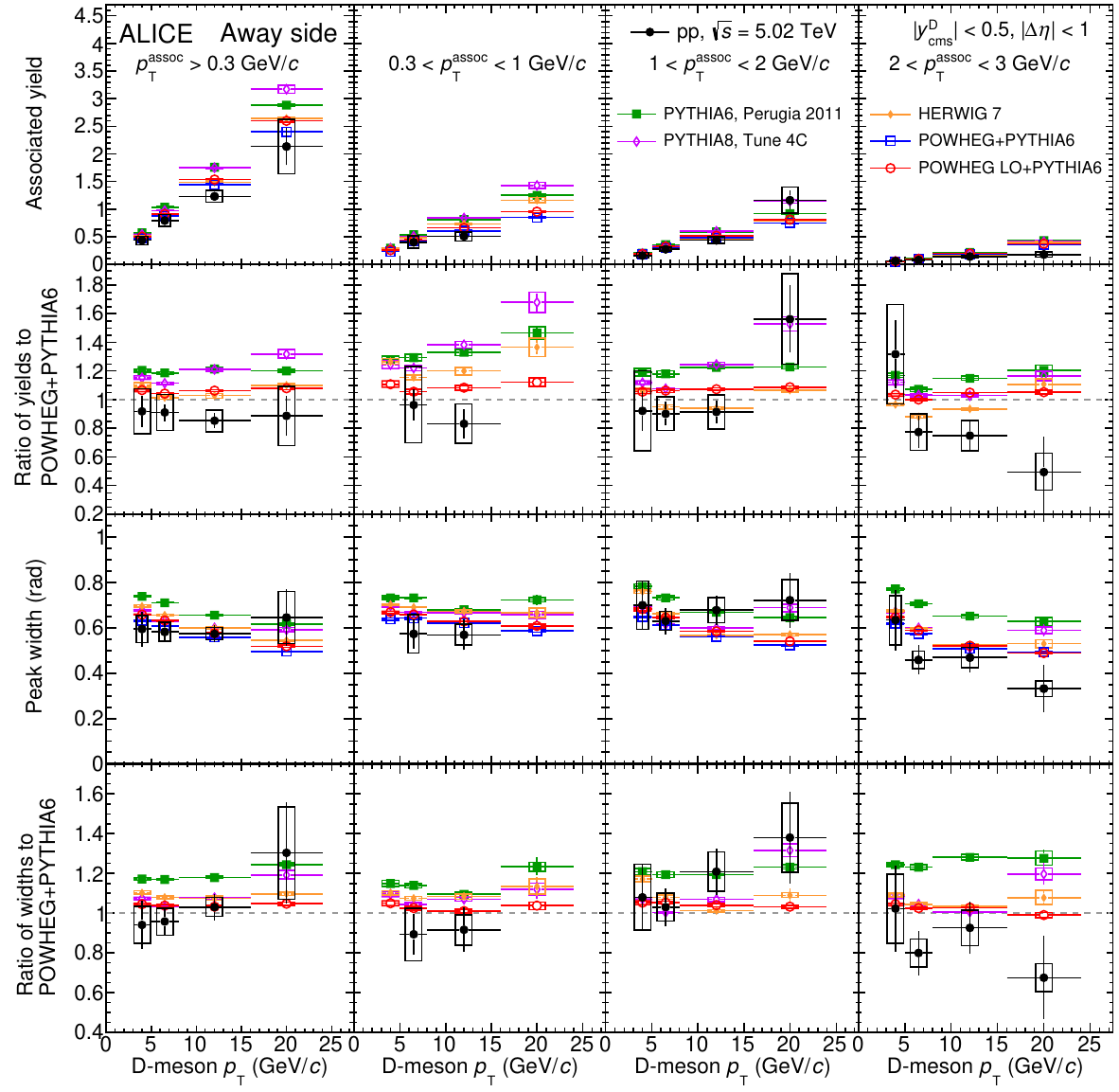}
    \end{center}
    \caption{Measurements of away-side associated peak yields (top row) and widths (third row down) in \pp collisions at $\s = 5.02$~\TeV, compared to predictions by the PYTHIA, POWHEG+PYTHIA6, POWHEG LO+PYTHIA6, and HERWIG event generators with various configurations (see text for details). The ratios of yield (width) values with respect to the predictions by POWHEG+PYTHIA6 are shown in the second (fourth) row down. Results are presented as a function of the D-meson $\pt$, for $\ptass > 0.3$~\GeVc, $0.3 < \ptass < 1$~\GeVc, $1 < \ptass < 2$~\GeVc, and $2 < \ptass < 3$~\GeVc (from left to right). Statistical and systematic uncertainties are shown as vertical error bars and boxes, respectively.}
    \label{fig:ppvsModels_AS}
\end{figure}

\begin{sloppypar}
The same comparison of model expectations to data is shown for the away-side peak yields and widths in Fig.~\ref{fig:ppvsModels_AS}.
POWHEG+PYTHIA6 gives the smallest away-side yields, with about 5\%--10\% smaller values than POWHEG LO+PYTHIA6 predictions. As for the near-side peak yields, this difference could be ascribed to a different contribution from back-to-back topologies of charm-quark pair production.
PYTHIA8 and PYTHIA6 yields are rather similar, and systematically larger than POWHEG LO+PYTHIA6 expectations. HERWIG predicts similar yields as POWHEG LO+PYTHIA6 for the integrated $\ptass$ range (with larger values for $0.3 < \ptass < 1$~\GeVc and smaller values for $\ptass > 1$~\GeVc).
The best description of the away-side yields is provided, as in the near-side peak case, by POWHEG+PYTHIA6 and POWHEG LO+PYTHIA6 over the whole kinematic range, as well as by HERWIG for $\ptass > 1$~\GeVc.
As observed for the near-side peak case, the PYTHIA8 and PYTHIA6 expectations tend to overpredict away-side yields in the majority of the transverse-momentum intervals studied. For $\ptass > 0.3$~\GeVc, about a $2\sigma$ difference with respect to the data is present, over the whole $\ptD$ interval studied.
The largest values of the away-side peak width, in particular for large values of $\ptass$, are given by the PYTHIA6 event generator, which tends to systematically overpredict the data points. The predictions from the other models, all in agreement with data, are very similar, with POWHEG+PYTHIA6 being in general the lowest of them.
However, the precision of measurements for this observable prevents from discerning the model that best describes the data.
\end{sloppypar}

\begin{figure}[tb]
    \begin{center}
    \includegraphics[width = 0.99\textwidth]{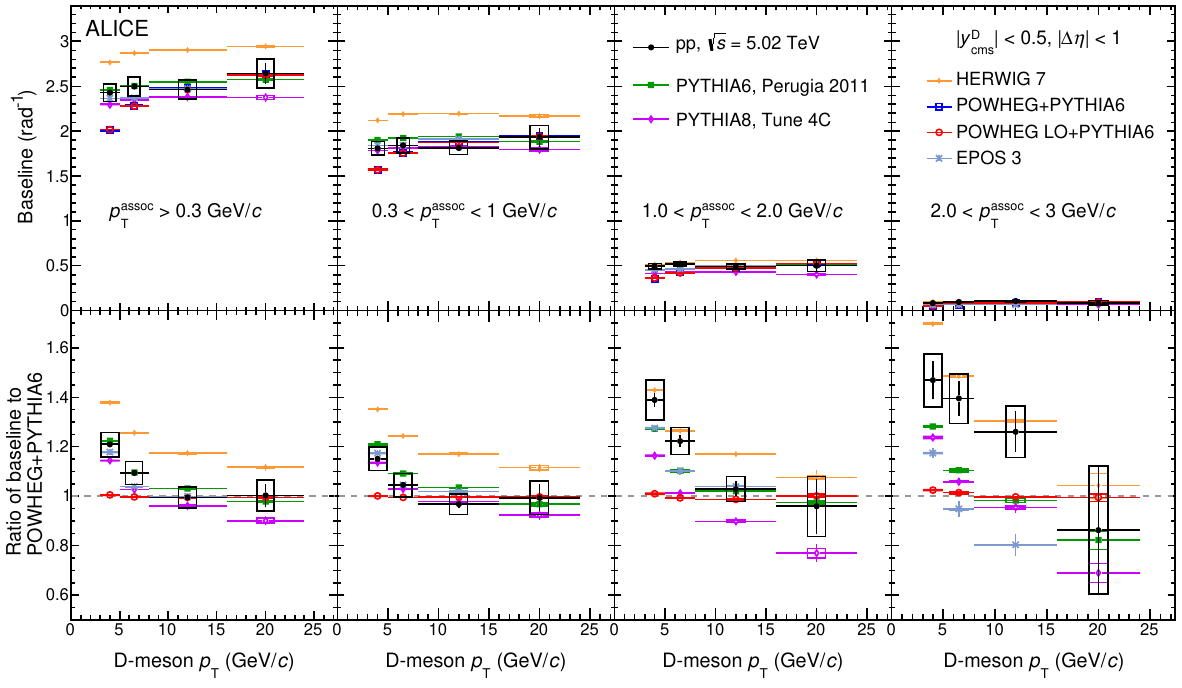}
    \end{center}
    \caption{Measurements of azimuthal-correlation function baseline height in \pp collisions at $\s = 5.02$ \TeV, compared to predictions by the PYTHIA, POWHEG+PYTHIA6, POWHEG LO+PYTHIA6, HERWIG, and EPOS 3 event generators with various configurations in the top row (see text for details). The ratios of baselines with respect to predictions by POWHEG+PYTHIA6 are shown in the bottom row. Results are presented as a function of the D-meson $\pt$, for $\ptass > 0.3$~\GeVc, $0.3 < \ptass < 1$~\GeVc, $1 < \ptass < 2$~\GeVc, and $2 < \ptass < 3$~\GeVc (from left to right). Statistical and systematic uncertainties are shown as vertical error bars and boxes, respectively.}
    \label{fig:ppvsModels_Basel}
\end{figure}

Figure~\ref{fig:ppvsModels_Basel} shows the baseline values of the measured azimuthal-correlation functions and compares them to predictions from the event generators.
The measured baseline values decrease with increasing $\pt$ of the associated particle, which is expected as the transverse-momentum distribution of associated particles in \pp collisions at $\s = 5.02$~\TeV peaks at few hundred \MeVc~\cite{Acharya:2018qsh}.
From the data it cannot be concluded whether the baseline is flat or slightly increasing as a function of D-meson $\pt$. A mildly increasing trend with $\ptD$ is predicted by the event generators. However, POWHEG+PYTHIA6 and POWHEG LO+PYTHIA6 predict a larger increase than HERWIG, EPOS 3 and PYTHIA.
The same baseline values are obtained by POWHEG+PYTHIA6 and POWHEG LO+PYTHIA6 for all the kinematic ranges. This is not trivial, due to the different treatment of next-to-leading order contributions to charm production, which can populate the transverse region of the correlation function and, hence, affect the baseline value.
The best description of the results, for low values of the associated particle $\pt$, is provided by PYTHIA6, PYTHIA8, and EPOS, while HERWIG overestimates the values by about 15\% over the whole $\ptD$ range and POWHEG+PYTHIA6 underpredicts them by 20\% at low $\ptD$.
For $\ptass > 1$~\GeVc HERWIG gives the closest description of the baseline. PYTHIA6, PYTHIA8, EPOS 3, and POWHEG+PYTHIA6 tend to underpredict data values, with the first three models catching well the $\ptD$ dependence, while POWHEG+PYTHIA6 also predicting a different behaviour against $\ptD$.

\begin{sloppypar}
The baseline-subtracted azimuthal-correlation functions of D mesons with associated particles measured in \pPb collisions were compared to simulations from PYTHIA6, PYTHIA8, and POWHEG+PYTHIA6 event generators. The only modifications of the configuration of these models with respect to that used in \pp collisions consisted of a rapidity shift of the centre-of-mass system and, for POWHEG+PYTHIA6, a nuclear correction for the parton distribution functions~\cite{Eskola:2009uj}, which induced negligible effects on the model expectations.
The comparison between these models and the results from \pPb collision yielded very similar conclusions as those discussed for \pp collisions, not only in terms of an overall agreement between data and models, but also for the differences previously mentioned for specific observables and kinematic ranges. This was expected, given the overall agreement of measurements in the two collision systems as discussed in Sec.~\ref{subsec:ppvspPb}, where additional cold-nuclear-matter effects, not included in the models, could also be present.
\end{sloppypar}

\clearpage

\section{Summary}
\label{sec:Summary}

Measurements of azimuthal-correlation functions of $\Dzero$, $\Dstar$, and $\Dplus$ mesons with charged particles in \pp collisions at $\s = 5.02$~\TeV and \pPb collisions at $\snn = 5.02$ \TeV were reported. The results obtained have statistical and systematic uncertainties smaller by a factor of about 2-3 than those reported in our previous paper~\cite{ALICE:2016clc} for the common $\pt$ ranges, and extend the \pt coverage both for the trigger and associated particles, allowing for a more differential study of the correlation function and charm-jet properties.

After subtracting the baseline, the correlation functions, along with the values and transverse momentum evolution of the near- and away-side peaks, are found to be consistent in \pp and \pPb collisions, in all the kinematic ranges addressed. This suggests that the fragmentation and hadronisation of charm quarks is not strongly influenced by cold-nuclear-matter effects, complementing what was observed in previous measurements~\cite{Adam:2016mkz,Adam:2015qda,Acharya:2019mno} that suggested a small impact from cold-nuclear-matter effects on D-meson production in the $\pt$ region covered by our measurement.

The analysis in \pPb collisions was also performed in the 0--20\%, 20--60\%, and 60--100\% centrality intervals, in order to study the possible modifications of the charm fragmentation as a function of the event centrality.
The same correlation pattern, along with similar values and $\pt$ evolution of the near-side peak observables were found for the three centrality ranges, within large experimental uncertainties.

The baseline-subtracted correlation functions and the near- and away-side peak yields and widths measured by ALICE in \pp collisions were compared to predictions by several event generators, with different modelling of charm production, parton showering, and hadronisation.
In general, the models describe well the main features of the correlation functions. POWHEG+PYTHIA6 provides the best description to experimental data of near- and away-side yields.
PYTHIA8 tends to overestimate the away-side peak yields, while providing a good description of the near-side peak yields and of the widths of both peaks.
Overall, PYTHIA6 is more distant from data than PYTHIA8, although in general it is consistent with the measurements.
HERWIG largely underestimates the near-side peak yields for $\ptass > 1$~\GeVc, while it describes reasonably well the data at lower $\ptass$, and provides a good description of the away-side peak features.
Finally, EPOS 3 provides a higher near-side peak and qualitatively underestimates the away-side peak.
Similar conclusions were obtained when comparing results in \pPb collisions to predictions from the models available in this collision system.

The agreement between data and model expectations suggests that charm-quark production, fragmentation and hadronisation processes, as implemented in POWHEG+PYTHIA6 and PYTHIA8, provide an overall satisfactory description of the measured correlation functions. Therefore, in view of future analyses in \PbPb collisions, these models constitute a valid theoretical baseline for interpreting
possible modifications of charm-jet properties and thus of the near-side correlation peak induced by the interactions of charm quarks with the quark--gluon plasma constituents.
The same argument holds for the modifications of the whole correlation function, whose characterisation can provide a deeper understanding of heavy-quark dynamics inside the QGP medium.
In addition, with the increased precision compared to previous measurement, and being at the same centre-of-mass energy of the available \PbPb collision samples, these results constitute the reference for measurements in that collision system.


\newenvironment{acknowledgement}{\relax}{\relax}
\begin{acknowledgement}
\section*{Acknowledgements}

The ALICE Collaboration would like to thank all its engineers and technicians for their invaluable contributions to the construction of the experiment and the CERN accelerator teams for the outstanding performance of the LHC complex.
The ALICE Collaboration gratefully acknowledges the resources and support provided by all Grid centres and the Worldwide LHC Computing Grid (WLCG) collaboration.
The ALICE Collaboration acknowledges the following funding agencies for their support in building and running the ALICE detector:
A. I. Alikhanyan National Science Laboratory (Yerevan Physics Institute) Foundation (ANSL), State Committee of Science and World Federation of Scientists (WFS), Armenia;
Austrian Academy of Sciences, Austrian Science Fund (FWF): [M 2467-N36] and Nationalstiftung f\"{u}r Forschung, Technologie und Entwicklung, Austria;
Ministry of Communications and High Technologies, National Nuclear Research Center, Azerbaijan;
Conselho Nacional de Desenvolvimento Cient\'{\i}fico e Tecnol\'{o}gico (CNPq), Financiadora de Estudos e Projetos (Finep), Funda\c{c}\~{a}o de Amparo \`{a} Pesquisa do Estado de S\~{a}o Paulo (FAPESP) and Universidade Federal do Rio Grande do Sul (UFRGS), Brazil;
Ministry of Education of China (MOEC) , Ministry of Science \& Technology of China (MSTC) and National Natural Science Foundation of China (NSFC), China;
Ministry of Science and Education and Croatian Science Foundation, Croatia;
Centro de Aplicaciones Tecnol\'{o}gicas y Desarrollo Nuclear (CEADEN), Cubaenerg\'{\i}a, Cuba;
Ministry of Education, Youth and Sports of the Czech Republic, Czech Republic;
The Danish Council for Independent Research | Natural Sciences, the VILLUM FONDEN and Danish National Research Foundation (DNRF), Denmark;
Helsinki Institute of Physics (HIP), Finland;
Commissariat \`{a} l'Energie Atomique (CEA), Institut National de Physique Nucl\'{e}aire et de Physique des Particules (IN2P3) and Centre National de la Recherche Scientifique (CNRS) and R\'{e}gion des  Pays de la Loire, France;
Bundesministerium f\"{u}r Bildung und Forschung (BMBF) and GSI Helmholtzzentrum f\"{u}r Schwerionenforschung GmbH, Germany;
General Secretariat for Research and Technology, Ministry of Education, Research and Religions, Greece;
National Research, Development and Innovation Office, Hungary;
Department of Atomic Energy Government of India (DAE), Department of Science and Technology, Government of India (DST), University Grants Commission, Government of India (UGC) and Council of Scientific and Industrial Research (CSIR), India;
Indonesian Institute of Science, Indonesia;
Centro Fermi - Museo Storico della Fisica e Centro Studi e Ricerche Enrico Fermi and Istituto Nazionale di Fisica Nucleare (INFN), Italy;
Institute for Innovative Science and Technology , Nagasaki Institute of Applied Science (IIST), Japanese Ministry of Education, Culture, Sports, Science and Technology (MEXT) and Japan Society for the Promotion of Science (JSPS) KAKENHI, Japan;
Consejo Nacional de Ciencia (CONACYT) y Tecnolog\'{i}a, through Fondo de Cooperaci\'{o}n Internacional en Ciencia y Tecnolog\'{i}a (FONCICYT) and Direcci\'{o}n General de Asuntos del Personal Academico (DGAPA), Mexico;
Nederlandse Organisatie voor Wetenschappelijk Onderzoek (NWO), Netherlands;
The Research Council of Norway, Norway;
Commission on Science and Technology for Sustainable Development in the South (COMSATS), Pakistan;
Pontificia Universidad Cat\'{o}lica del Per\'{u}, Peru;
Ministry of Science and Higher Education and National Science Centre, Poland;
Korea Institute of Science and Technology Information and National Research Foundation of Korea (NRF), Republic of Korea;
Ministry of Education and Scientific Research, Institute of Atomic Physics and Ministry of Research and Innovation and Institute of Atomic Physics, Romania;
Joint Institute for Nuclear Research (JINR), Ministry of Education and Science of the Russian Federation, National Research Centre Kurchatov Institute, Russian Science Foundation and Russian Foundation for Basic Research, Russia;
Ministry of Education, Science, Research and Sport of the Slovak Republic, Slovakia;
National Research Foundation of South Africa, South Africa;
Swedish Research Council (VR) and Knut \& Alice Wallenberg Foundation (KAW), Sweden;
European Organization for Nuclear Research, Switzerland;
Suranaree University of Technology (SUT), National Science and Technology Development Agency (NSDTA) and Office of the Higher Education Commission under NRU project of Thailand, Thailand;
Turkish Atomic Energy Agency (TAEK), Turkey;
National Academy of  Sciences of Ukraine, Ukraine;
Science and Technology Facilities Council (STFC), United Kingdom;
National Science Foundation of the United States of America (NSF) and United States Department of Energy, Office of Nuclear Physics (DOE NP), United States of America.
\end{acknowledgement}

\bibliographystyle{utphys}   
\bibliography{bibliography}

\newpage
\appendix

%
%

\section{The ALICE Collaboration}
\label{app:collab}

\begingroup
\small
\begin{flushleft}
S.~Acharya\Irefn{org141}\And 
D.~Adamov\'{a}\Irefn{org94}\And 
A.~Adler\Irefn{org74}\And 
J.~Adolfsson\Irefn{org80}\And 
M.M.~Aggarwal\Irefn{org99}\And 
G.~Aglieri Rinella\Irefn{org33}\And 
M.~Agnello\Irefn{org30}\And 
N.~Agrawal\Irefn{org10}\textsuperscript{,}\Irefn{org53}\And 
Z.~Ahammed\Irefn{org141}\And 
S.~Ahmad\Irefn{org16}
% ...
\And
\renewcommand\labelenumi{\textsuperscript{\theenumi}~}

\section*{Affiliation notes}
\renewcommand\theenumi{\roman{enumi}}
\begin{Authlist}
\item \Adef{org*}Deceased
\item \Adef{orgI}Dipartimento DET del Politecnico di Torino, Turin, Italy
\item \Adef{orgII}M.V. Lomonosov Moscow State University, D.V. Skobeltsyn Institute of Nuclear, Physics, Moscow, Russia
\item \Adef{orgIII}Department of Applied Physics, Aligarh Muslim University, Aligarh, India
\item \Adef{orgIV}Institute of Theoretical Physics, University of Wroclaw, Poland
\end{Authlist}

\section*{Collaboration Institutes}
\renewcommand\theenumi{\arabic{enumi}~}
\begin{Authlist}
\item \Idef{org1}A.I. Alikhanyan National Science Laboratory (Yerevan Physics Institute) Foundation, Yerevan, Armenia
\item \Idef{org2}Bogolyubov Institute for Theoretical Physics, National Academy of Sciences of Ukraine, Kiev, Ukraine
\item \Idef{org3}Bose Institute, Department of Physics  and Centre for Astroparticle Physics and Space Science (CAPSS), Kolkata, India
\item \Idef{org4}Budker Institute for Nuclear Physics, Novosibirsk, Russia
\item \Idef{org5}California Polytechnic State University, San Luis Obispo, California, United States
\item \Idef{org6}Central China Normal University, Wuhan, China
\item \Idef{org7}Centre de Calcul de l'IN2P3, Villeurbanne, Lyon, France
\item \Idef{org8}Centro de Aplicaciones Tecnol\'{o}gicas y Desarrollo Nuclear (CEADEN), Havana, Cuba
\item \Idef{org9}Centro de Investigaci\'{o}n y de Estudios Avanzados (CINVESTAV), Mexico City and M\'{e}rida, Mexico
\item \Idef{org10}Centro Fermi - Museo Storico della Fisica e Centro Studi e Ricerche ``Enrico Fermi', Rome, Italy
\item \Idef{org11}Chicago State University, Chicago, Illinois, United States
\item \Idef{org12}China Institute of Atomic Energy, Beijing, China
\item \Idef{org13}Comenius University Bratislava, Faculty of Mathematics, Physics and Informatics, Bratislava, Slovakia
\item \Idef{org14}COMSATS University Islamabad, Islamabad, Pakistan
\item \Idef{org15}Creighton University, Omaha, Nebraska, United States
\item \Idef{org16}Department of Physics, Aligarh Muslim University, Aligarh, India
\item \Idef{org17}Department of Physics, Pusan National University, Pusan, Republic of Korea
\item \Idef{org18}Department of Physics, Sejong University, Seoul, Republic of Korea
\item \Idef{org19}Department of Physics, University of California, Berkeley, California, United States
\item \Idef{org20}Department of Physics, University of Oslo, Oslo, Norway
\item \Idef{org21}Department of Physics and Technology, University of Bergen, Bergen, Norway
\item \Idef{org22}Dipartimento di Fisica dell'Universit\`{a} 'La Sapienza' and Sezione INFN, Rome, Italy
\item \Idef{org23}Dipartimento di Fisica dell'Universit\`{a} and Sezione INFN, Cagliari, Italy
\item \Idef{org24}Dipartimento di Fisica dell'Universit\`{a} and Sezione INFN, Trieste, Italy
\item \Idef{org25}Dipartimento di Fisica dell'Universit\`{a} and Sezione INFN, Turin, Italy
\item \Idef{org26}Dipartimento di Fisica e Astronomia dell'Universit\`{a} and Sezione INFN, Bologna, Italy
\item \Idef{org27}Dipartimento di Fisica e Astronomia dell'Universit\`{a} and Sezione INFN, Catania, Italy
\item \Idef{org28}Dipartimento di Fisica e Astronomia dell'Universit\`{a} and Sezione INFN, Padova, Italy
\item \Idef{org29}Dipartimento di Fisica `E.R.~Caianiello' dell'Universit\`{a} and Gruppo Collegato INFN, Salerno, Italy
\item \Idef{org30}Dipartimento DISAT del Politecnico and Sezione INFN, Turin, Italy
\item \Idef{org31}Dipartimento di Scienze e Innovazione Tecnologica dell'Universit\`{a} del Piemonte Orientale and INFN Sezione di Torino, Alessandria, Italy
\item \Idef{org32}Dipartimento Interateneo di Fisica `M.~Merlin' and Sezione INFN, Bari, Italy
\item \Idef{org33}European Organization for Nuclear Research (CERN), Geneva, Switzerland
\item \Idef{org34}Faculty of Electrical Engineering, Mechanical Engineering and Naval Architecture, University of Split, Split, Croatia
\item \Idef{org35}Faculty of Engineering and Science, Western Norway University of Applied Sciences, Bergen, Norway
\item \Idef{org36}Faculty of Nuclear Sciences and Physical Engineering, Czech Technical University in Prague, Prague, Czech Republic
\item \Idef{org37}Faculty of Science, P.J.~\v{S}af\'{a}rik University, Ko\v{s}ice, Slovakia
\item \Idef{org38}Frankfurt Institute for Advanced Studies, Johann Wolfgang Goethe-Universit\"{a}t Frankfurt, Frankfurt, Germany
\item \Idef{org39}Fudan University, Shanghai, China
\item \Idef{org40}Gangneung-Wonju National University, Gangneung, Republic of Korea
\item \Idef{org41}Gauhati University, Department of Physics, Guwahati, India
\item \Idef{org42}Helmholtz-Institut f\"{u}r Strahlen- und Kernphysik, Rheinische Friedrich-Wilhelms-Universit\"{a}t Bonn, Bonn, Germany
\item \Idef{org43}Helsinki Institute of Physics (HIP), Helsinki, Finland
\item \Idef{org44}High Energy Physics Group,  Universidad Aut\'{o}noma de Puebla, Puebla, Mexico
\item \Idef{org45}Hiroshima University, Hiroshima, Japan
\item \Idef{org46}Hochschule Worms, Zentrum  f\"{u}r Technologietransfer und Telekommunikation (ZTT), Worms, Germany
\item \Idef{org47}Horia Hulubei National Institute of Physics and Nuclear Engineering, Bucharest, Romania
\item \Idef{org48}Indian Institute of Technology Bombay (IIT), Mumbai, India
\item \Idef{org49}Indian Institute of Technology Indore, Indore, India
\item \Idef{org50}Indonesian Institute of Sciences, Jakarta, Indonesia
\item \Idef{org51}INFN, Laboratori Nazionali di Frascati, Frascati, Italy
\item \Idef{org52}INFN, Sezione di Bari, Bari, Italy
\item \Idef{org53}INFN, Sezione di Bologna, Bologna, Italy
\item \Idef{org54}INFN, Sezione di Cagliari, Cagliari, Italy
\item \Idef{org55}INFN, Sezione di Catania, Catania, Italy
\item \Idef{org56}INFN, Sezione di Padova, Padova, Italy
\item \Idef{org57}INFN, Sezione di Roma, Rome, Italy
\item \Idef{org58}INFN, Sezione di Torino, Turin, Italy
\item \Idef{org59}INFN, Sezione di Trieste, Trieste, Italy
\item \Idef{org60}Inha University, Incheon, Republic of Korea
\item \Idef{org61}Institut de Physique Nucl\'{e}aire d'Orsay (IPNO), Institut National de Physique Nucl\'{e}aire et de Physique des Particules (IN2P3/CNRS), Universit\'{e} de Paris-Sud, Universit\'{e} Paris-Saclay, Orsay, France
\item \Idef{org62}Institute for Nuclear Research, Academy of Sciences, Moscow, Russia
\item \Idef{org63}Institute for Subatomic Physics, Utrecht University/Nikhef, Utrecht, Netherlands
\item \Idef{org64}Institute of Experimental Physics, Slovak Academy of Sciences, Ko\v{s}ice, Slovakia
\item \Idef{org65}Institute of Physics, Homi Bhabha National Institute, Bhubaneswar, India
\item \Idef{org66}Institute of Physics of the Czech Academy of Sciences, Prague, Czech Republic
\item \Idef{org67}Institute of Space Science (ISS), Bucharest, Romania
\item \Idef{org68}Institut f\"{u}r Kernphysik, Johann Wolfgang Goethe-Universit\"{a}t Frankfurt, Frankfurt, Germany
\item \Idef{org69}Instituto de Ciencias Nucleares, Universidad Nacional Aut\'{o}noma de M\'{e}xico, Mexico City, Mexico
\item \Idef{org70}Instituto de F\'{i}sica, Universidade Federal do Rio Grande do Sul (UFRGS), Porto Alegre, Brazil
\item \Idef{org71}Instituto de F\'{\i}sica, Universidad Nacional Aut\'{o}noma de M\'{e}xico, Mexico City, Mexico
\item \Idef{org72}iThemba LABS, National Research Foundation, Somerset West, South Africa
\item \Idef{org73}Jeonbuk National University, Jeonju, Republic of Korea
\item \Idef{org74}Johann-Wolfgang-Goethe Universit\"{a}t Frankfurt Institut f\"{u}r Informatik, Fachbereich Informatik und Mathematik, Frankfurt, Germany
\item \Idef{org75}Joint Institute for Nuclear Research (JINR), Dubna, Russia
\item \Idef{org76}Korea Institute of Science and Technology Information, Daejeon, Republic of Korea
\item \Idef{org77}KTO Karatay University, Konya, Turkey
\item \Idef{org78}Laboratoire de Physique Subatomique et de Cosmologie, Universit\'{e} Grenoble-Alpes, CNRS-IN2P3, Grenoble, France
\item \Idef{org79}Lawrence Berkeley National Laboratory, Berkeley, California, United States
\item \Idef{org80}Lund University Department of Physics, Division of Particle Physics, Lund, Sweden
\item \Idef{org81}Nagasaki Institute of Applied Science, Nagasaki, Japan
\item \Idef{org82}Nara Women{'}s University (NWU), Nara, Japan
\item \Idef{org83}National and Kapodistrian University of Athens, School of Science, Department of Physics , Athens, Greece
\item \Idef{org84}National Centre for Nuclear Research, Warsaw, Poland
\item \Idef{org85}National Institute of Science Education and Research, Homi Bhabha National Institute, Jatni, India
\item \Idef{org86}National Nuclear Research Center, Baku, Azerbaijan
\item \Idef{org87}National Research Centre Kurchatov Institute, Moscow, Russia
\item \Idef{org88}Niels Bohr Institute, University of Copenhagen, Copenhagen, Denmark
\item \Idef{org89}Nikhef, National institute for subatomic physics, Amsterdam, Netherlands
\item \Idef{org90}NRC Kurchatov Institute IHEP, Protvino, Russia
\item \Idef{org91}NRC «Kurchatov Institute»  - ITEP, Moscow, Russia
\item \Idef{org92}NRNU Moscow Engineering Physics Institute, Moscow, Russia
\item \Idef{org93}Nuclear Physics Group, STFC Daresbury Laboratory, Daresbury, United Kingdom
\item \Idef{org94}Nuclear Physics Institute of the Czech Academy of Sciences, \v{R}e\v{z} u Prahy, Czech Republic
\item \Idef{org95}Oak Ridge National Laboratory, Oak Ridge, Tennessee, United States
\item \Idef{org96}Ohio State University, Columbus, Ohio, United States
\item \Idef{org97}Petersburg Nuclear Physics Institute, Gatchina, Russia
\item \Idef{org98}Physics department, Faculty of science, University of Zagreb, Zagreb, Croatia
\item \Idef{org99}Physics Department, Panjab University, Chandigarh, India
\item \Idef{org100}Physics Department, University of Jammu, Jammu, India
\item \Idef{org101}Physics Department, University of Rajasthan, Jaipur, India
\item \Idef{org102}Physikalisches Institut, Eberhard-Karls-Universit\"{a}t T\"{u}bingen, T\"{u}bingen, Germany
\item \Idef{org103}Physikalisches Institut, Ruprecht-Karls-Universit\"{a}t Heidelberg, Heidelberg, Germany
\item \Idef{org104}Physik Department, Technische Universit\"{a}t M\"{u}nchen, Munich, Germany
\item \Idef{org105}Politecnico di Bari, Bari, Italy
\item \Idef{org106}Research Division and ExtreMe Matter Institute EMMI, GSI Helmholtzzentrum f\"ur Schwerionenforschung GmbH, Darmstadt, Germany
\item \Idef{org107}Rudjer Bo\v{s}kovi\'{c} Institute, Zagreb, Croatia
\item \Idef{org108}Russian Federal Nuclear Center (VNIIEF), Sarov, Russia
\item \Idef{org109}Saha Institute of Nuclear Physics, Homi Bhabha National Institute, Kolkata, India
\item \Idef{org110}School of Physics and Astronomy, University of Birmingham, Birmingham, United Kingdom
\item \Idef{org111}Secci\'{o}n F\'{\i}sica, Departamento de Ciencias, Pontificia Universidad Cat\'{o}lica del Per\'{u}, Lima, Peru
\item \Idef{org112}St. Petersburg State University, St. Petersburg, Russia
\item \Idef{org113}Stefan Meyer Institut f\"{u}r Subatomare Physik (SMI), Vienna, Austria
\item \Idef{org114}SUBATECH, IMT Atlantique, Universit\'{e} de Nantes, CNRS-IN2P3, Nantes, France
\item \Idef{org115}Suranaree University of Technology, Nakhon Ratchasima, Thailand
\item \Idef{org116}Technical University of Ko\v{s}ice, Ko\v{s}ice, Slovakia
\item \Idef{org117}Technische Universit\"{a}t M\"{u}nchen, Excellence Cluster 'Universe', Munich, Germany
\item \Idef{org118}The Henryk Niewodniczanski Institute of Nuclear Physics, Polish Academy of Sciences, Cracow, Poland
\item \Idef{org119}The University of Texas at Austin, Austin, Texas, United States
\item \Idef{org120}Universidad Aut\'{o}noma de Sinaloa, Culiac\'{a}n, Mexico
\item \Idef{org121}Universidade de S\~{a}o Paulo (USP), S\~{a}o Paulo, Brazil
\item \Idef{org122}Universidade Estadual de Campinas (UNICAMP), Campinas, Brazil
\item \Idef{org123}Universidade Federal do ABC, Santo Andre, Brazil
\item \Idef{org124}University of Cape Town, Cape Town, South Africa
\item \Idef{org125}University of Houston, Houston, Texas, United States
\item \Idef{org126}University of Jyv\"{a}skyl\"{a}, Jyv\"{a}skyl\"{a}, Finland
\item \Idef{org127}University of Liverpool, Liverpool, United Kingdom
\item \Idef{org128}University of Science and Technology of China, Hefei, China
\item \Idef{org129}University of South-Eastern Norway, Tonsberg, Norway
\item \Idef{org130}University of Tennessee, Knoxville, Tennessee, United States
\item \Idef{org131}University of the Witwatersrand, Johannesburg, South Africa
\item \Idef{org132}University of Tokyo, Tokyo, Japan
\item \Idef{org133}University of Tsukuba, Tsukuba, Japan
\item \Idef{org134}Universit\'{e} Clermont Auvergne, CNRS/IN2P3, LPC, Clermont-Ferrand, France
\item \Idef{org135}Universit\'{e} de Lyon, Universit\'{e} Lyon 1, CNRS/IN2P3, IPN-Lyon, Villeurbanne, Lyon, France
\item \Idef{org136}Universit\'{e} de Strasbourg, CNRS, IPHC UMR 7178, F-67000 Strasbourg, France, Strasbourg, France
\item \Idef{org137}Universit\'{e} Paris-Saclay Centre d'Etudes de Saclay (CEA), IRFU, D\'{e}partment de Physique Nucl\'{e}aire (DPhN), Saclay, France
\item \Idef{org138}Universit\`{a} degli Studi di Foggia, Foggia, Italy
\item \Idef{org139}Universit\`{a} degli Studi di Pavia, Pavia, Italy
\item \Idef{org140}Universit\`{a} di Brescia, Brescia, Italy
\item \Idef{org141}Variable Energy Cyclotron Centre, Homi Bhabha National Institute, Kolkata, India
\item \Idef{org142}Warsaw University of Technology, Warsaw, Poland
\item \Idef{org143}Wayne State University, Detroit, Michigan, United States
\item \Idef{org144}Westf\"{a}lische Wilhelms-Universit\"{a}t M\"{u}nster, Institut f\"{u}r Kernphysik, M\"{u}nster, Germany
\item \Idef{org145}Wigner Research Centre for Physics, Budapest, Hungary
\item \Idef{org146}Yale University, New Haven, Connecticut, United States
\item \Idef{org147}Yonsei University, Seoul, Republic of Korea
\end{Authlist}
\endgroup
  
\end{document}